\documentclass[a4paper]{aastex}
\usepackage{graphicx}
\usepackage{natbib}
\bibpunct{(}{)}{;}{a}{}{,}
\usepackage{amsmath}

\shorttitle{Warm Breeze in the heliosphere}
\shortauthors{M.A. Kubiak, M. Bzowski et al.}

\begin{document}

\title{Warm Breeze from the starboard bow: a new population of neutral helium in the heliosphere}
    
\author{M.A. Kubiak\altaffilmark{1}, M. Bzowski\altaffilmark{1}, J.M. Sok\'o\l\altaffilmark{1}, P. Swaczyna\altaffilmark{1}, S. Grzedzielski\altaffilmark{1}, D.B. Alexashov\altaffilmark{2,3}, V.V. Izmodenov\altaffilmark{2,3,4}, E. M\"{o}bius\altaffilmark{5}, T. Leonard\altaffilmark{5}, S.A. Fuselier\altaffilmark{6,7}, P. Wurz\altaffilmark{8}, D.J. McComas\altaffilmark{6,7}}

\altaffiltext{1}{Space Research Centre of the Polish Academy of Sciences, Warsaw, Poland}
\altaffiltext{2}{Space Research Institute (IKI) of the Russian Academy of Sciences, Moscow, Russia}
\altaffiltext{3}{Institute for Problems in Mechanics of the Russian Academy of Sciences, Moscow, Russia}
\altaffiltext{4}{Lomonosov Moscow State University, Moscow, Russia}
\altaffiltext{5}{ Space Research Center and Department of Physics, University of New Hampshire, Durham, NH USA}
\altaffiltext{5}{Southwest Research Institute, San Antonio, TX USA}
\altaffiltext{5}{The University of Texas at San Antonio, TX USA}
\altaffiltext{5}{Physics Institute, University of Bern, Bern, Switzerland}

\begin{abstract}
We investigate the signals from neutral helium atoms observed in situ from Earth orbit in 2010 by the Interstellar Boundary Explorer (IBEX). The full helium signal observed during the 2010 observation season can be explained as a superposition of pristine neutral interstellar He gas and an additional population of neutral helium that we call the Warm Breeze. The Warm Breeze is approximately two-fold slower and 2.5 times warmer than the primary interstellar He population, and its density in front of the heliosphere is $\sim$7\% that of the neutral interstellar helium. The inflow direction of the Warm Breeze differs by $\sim$19$^{\circ}$ from the inflow direction of interstellar gas. The Warm Breeze seems a long-term, perhaps permanent feature of the heliospheric environment. It has not been detected earlier because it is strongly ionized inside the heliosphere. This effect brings it below the threshold of detection via pickup ion and heliospheric backscatter glow observations, as well as by the direct sampling of GAS/Ulysses. We discuss possible sources for the Warm Breeze, including (1) the secondary population of interstellar helium, created via charge exchange and perhaps elastic scattering of neutral interstellar He atoms on interstellar He$^+$ ions in the outer heliosheath, or (2) a gust of interstellar He originating from a hypothetic wave train in the Local Interstellar Cloud. A secondary population is expected from models, but the characteristics of the Warm Breeze do not fully conform to modeling results. If, nevertheless, this is the explanation, IBEX-Lo observations of the Warm Breeze provide key insights into the physical state of plasma in the outer heliosheath. If the second hypothesis is true, the source is likely to be located within a few thousand of AU from the Sun, which is the propagation range of possible gusts of interstellar neutral helium with the Warm Breeze characteristics against dissipation via elastic scattering in the Local Cloud. Whatever the nature of the Warm Breeze, its discovery exposes a critical new feature of our heliospheric environment.

\end{abstract}


\section{Introduction}
Studies of the heliosphere and the surrounding interstellar medium have recently expanded tremendously owing to the new observation capabilities offered by the Interstellar Boundary Explorer mission (IBEX) \citep{mccomas_etal:09a}. After the GAS experiment on Ulysses \citep{witte_etal:92a}, the IBEX-Lo instrument \citep{fuselier_etal:09b} is the only other able to study the neutral component of interstellar matter surrounding the heliosphere via direct sampling of neutral interstellar (NIS) atoms. Following the detection of NIS He, O, Ne, and H by \citet{mobius_etal:09b}, \citet{bzowski_etal:12a} and \citet{mobius_etal:12a} concluded that the inflow of NIS gas on the heliosphere is slower by $\sim$ 3~km/s than inferred from Ulysses observations by \citet{witte:04} and comes from ecliptic longitude larger by $\sim~4\degr$. This discovery led \citet{mccomas_etal:12b}, as well as a number of other researchers \citep{benjaffel_ratkiewicz:12a, zank_etal:13a}, to confirm an earlier suggestion (e.g., \citet{gayley_etal:97, izmodenov_etal:09a}), supported by the Voyager 1 and Voyager 2 termination shock crossings at 84~AU and 94~AU, respectively, that a bow shock (BS) in front of the heliosphere may be absent, a posibility that had already been earlier envsioned in the modeling work \citep[e.g., ][]{mueller_etal:00}. The BS was expected in the earlier models (e.g., \citet{baranov_etal:70a, baranov_malama:93, pogorelov_semenov:97a, mueller_etal:00}). Review of all observations of NIS He flow through the heliosphere available from the beginning of space age led \citet{frisch_etal:13a} to realize that the direction of inflow of NIS He on the heliosphere may be changing at a mean rate of $\sim 0.17\degr$ per year.

\begin{figure}
\epsscale{.80}
\plotone{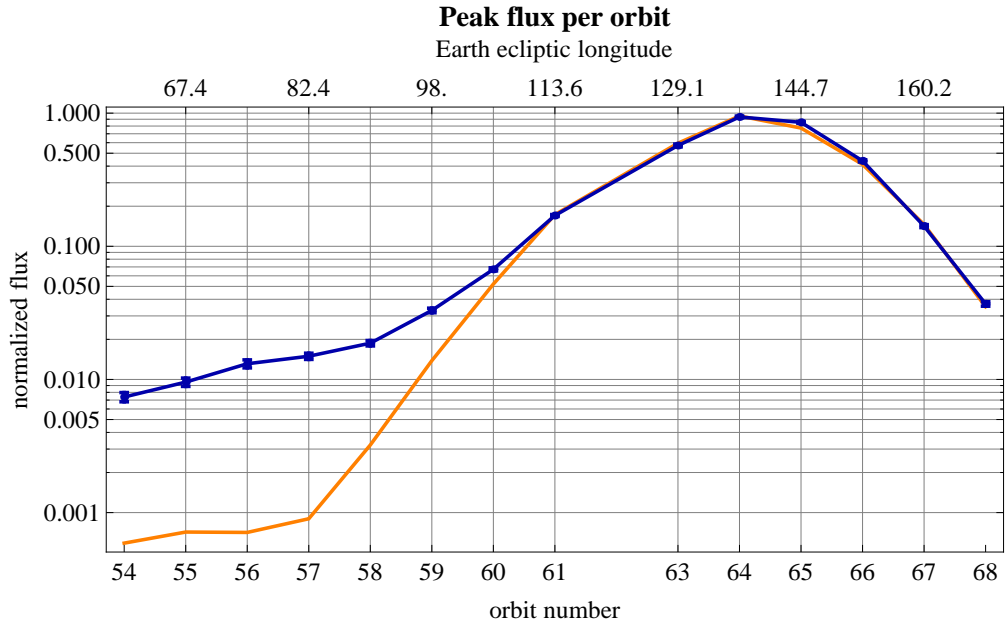}
\caption{Peak flux of neutral helium observed by IBEX on orbits 54 through 68 during the 2010 observation season (dark blue symbols with error bars), compared with the modeled peaks (orange) for the NIS He gas population for the flow parameters found by \citet{bzowski_etal:12a}. Both observed and model values are scaled to the respective values for orbit 64, when the season maximum flux was observed. The lower axis presents IBEX orbit numbers, the upper axis the corresponding ecliptic longitudes of Earth. }.\label{fig:fig_peakFlux}
\end{figure}

In addition to reporting the new inflow parameters of NIS He, \citet{bzowski_etal:12a} announced the discovery of a new population of NIS He gas, which they tentatively attributed to the secondary population of interstellar neutrals. Such a population is expected to form via charge exchange between pristine interstellar neutral gas and the interstellar plasma in the outer heliosheath region (i.e., beyond the heliopause). This population was seen as an excess of the observed signal over the one-component fit to the NIS He inflow (Fig.~\ref{fig:fig_peakFlux}) during the early orbits of the IBEX NIS gas observation seasons. Also, the one-component Maxwellian inflow could not explain a portion (not shown) of the observed signal with elevated wings on both sides of the signal from the primary population of NIS gas. In this paper we investigate these signals and their implications in more detail. 

\section{Observations}
\subsection{Data collection}

IBEX is a spin-stabilized spacecraft following a highly elliptical orbit around the Earth. The boresight of the IBEX-Lo instrument is perpendicular to the spin axis \citep{hlond_etal:12a}, which is adjusted at the beginning of each IBEX orbit to maintain it within $\sim$7$^{\circ}$ from the Sun. Interstellar atoms can be detected only when the IBEX-Lo aperture is looking into the flow, which happens during the first quarter of each year. The precise orientation of the IBEX spin axis during the observations used in this paper was presented by \citet{bzowski_etal:12a}. 

Details of NIS gas data acquisition by IBEX-Lo were presented by \citet{mobius_etal:12a}. IBEX-Lo measures energetic neutral atoms in eight wide, logarithmically spaced energy bands (steps), which are sequentially stepped over 64 spacecraft spins (which take approximately 16 minutes). The observations cover 360$^{\circ}$  of spin angle and are binned in 6$^{\circ}$ intervals. The effective field of view is approximately 7$^{\circ}$ FWHM in diameter and the FWHM width of the energy step is approximately 70\% of the central energy value. The atoms enter the instrument through a collimator, whose orientation, shape, and transmission function were presented by \citet{fuselier_etal:09b} and in Figs 2 and 3 in \citet{bzowski_etal:12a}. Having passed the collimator, the atoms hit a specially prepared conversion surface \citep{fuselier_etal:09b}, which is made of tetrahedral amorphous carbon and covered with a thin water layer, constantly renewed due to instrument outgassing.

The instrument does not directly measure NIS He atoms because He forms only metastable He$^-$ ion. However, NIS He atoms with energies above $\sim$10 -- 30 eV, impacting at the IBEX-Lo conversion surface, sputter H, O, and C atoms, some of which are in the form of negative ions that can be detected. The atoms with energies below an energy threshold are incapable of sputtering \citep{yamamura_tawara:96a}, but in the case of the IBEX conversion surface, the sputtering limit is likely reduced due to the water layer on the carbon surface \citep{taglauer:90a}. The effective energy threshold is formed as a combination of the sputtering threshold and effective probability of registration of negative ions as a function of their energy. Sputter products have lower energies than the parent atom and are observed over all IBEX-Lo energy steps below the incoming energy \citep{mobius_etal:12a}.

As calibrated before flight, the probability of generating a sputtering product in the energy range of Step 1 through 3 (with central energies 14.5, 28.5, and 55.5 eV, respectively) is a weak function of the impact energy above the sputtering threshold. Thus, for analysis of NIS He one can select any of energy steps 1 through 3. The signal that we interpret as due to the Warm Breeze is present in all three lowest energy channels of IBEX-Lo. However, the signal in energy step 1 includes NIS H during the later orbits \citep{saul_etal:12a}, and the signal collected in November -- January seems to be due to less energetic atoms than the primary population, which would lead to stronger efficiency variations in energy step 3. Therefore, we use IBEX-Lo energy step 2 throughout the analysis. 

\begin{figure}
\epsscale{1.0}
\plotone{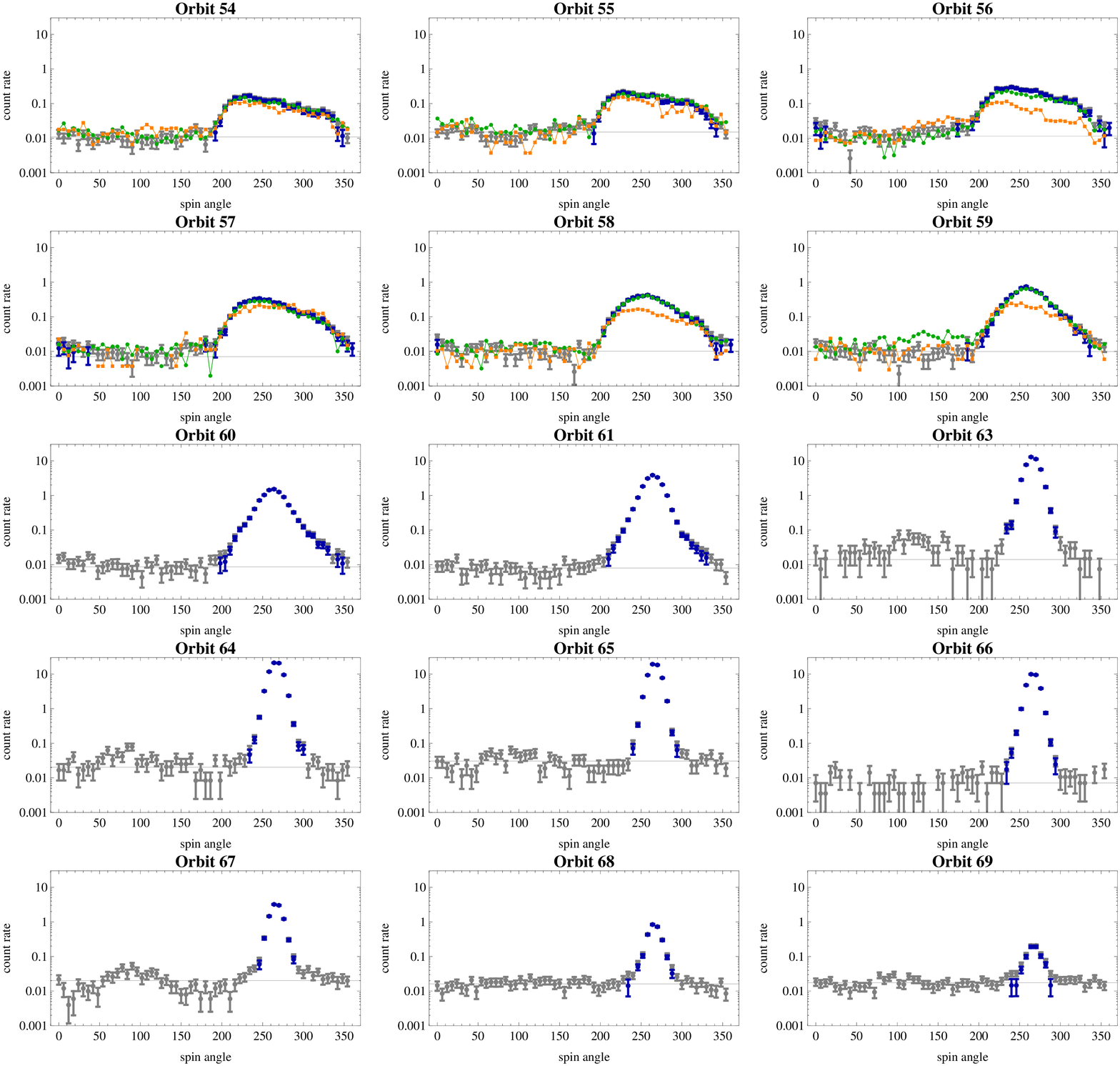}
\caption{Data used in the analysis. Gray dots with error bars represent count rates averaged over the lengths of good times intervals for individual orbits, obtained in the 2010 observation season. Shown are orbit-averaged count rates in (s bin)$^{-1}$ as a function of IBEX spin angle. The horizontal lines represent the adopted background levels for the orbits. Dark blue symbols represent the data actually adopted for the analysis, with background subtracted. Note that the vertical scale is logarithmic and owing to the very high signal to noise ratio the differences between the gray an blue dots may be hardly visible in the plot. Green and orange symbols represent orbit-averaged count rates from equivalent orbits from the 2011 and 2012 observation seasons, respectively, during  which the Warm Breeze population dominates. Orbit 62 is missing because of an unplanned reset of the on-board computer and resulting total data loss for this orbit \citep{mccomas_etal:12c}.}\label{fig:fig_dataPlot}
\end{figure}

The data for the detailed analysis were taken  between $\sim$60$^{\circ}$ and 170$^{\circ}$ ecliptic longitude, which corresponds to mid November 2009 through early April 2010, i.e., during the first season that covered the region relevant for the Warm Breeze after IBEX commissioning in late 2008 and early 2009. The counts from energy step 2 were integrated over the full duration of the NIS flow good times. The criteria for good times were discussed in detail by \citet{mobius_etal:12a}. The data we used are an extension of the subset analyzed by \citet{mobius_etal:12a} and \citet{bzowski_etal:12a}. As described by \citet{mobius_etal:12a}, the observed count rates were corrected for the limited throughput of direct events due to buffer and telemetry limitations for high count rates during the interstellar gas flow observations. For this correction, the filtered direct event data are accumulated in 6\degr~bins and normalized to 6\degr~resolution histograms that have been accumulated onboard before any of these throughput limitations. These small corrections vary in magnitude and are determined on a point by point basis. Necessarily, they are the largest for the highest count rates. While they are significant for the peak of the interstellar flow, they are very small for the Warm Breeze signal because of much lower count rates. Consequently, the contribution of these corrections to the Warm Breeze measurement uncertainties is practically negligible. 

The data used for the analysis are presented as dark blue symbols in Fig.~\ref{fig:fig_dataPlot}. They represent the count rate averaged over the observation time, specifically: total counts accumulated in a given bin, adjusted for the duty cycle of 60 angle bins and 8 energy steps. This quantity is directly proportional to the good time-average flux of NIS He impacting the spacecraft during a given orbit.

\subsection{Background removal and determination of count rate uncertainty}

The expected background sources in IBEX-Lo measurements, as well as methods used for their suppression, were presented in detail by \citet{wurz_etal:09a} and the in-flight performance is discussed by \citet{fuselier_etal:14a}. \citet{bzowski_etal:12a} used a subset of the selected data set with the signal to noise ratio of $\sim$100 or higher, and consequently did not need to seriously worry about the measurement background in the analysis. In contrast, here we focus on a portion of the data with a much lower signal to noise ratio and therefore have to analyze and subtract background. 

The data were thoroughly cleaned from all possible contamination. The portion of the signal we classified as background is predominantly due to genuine neutral atoms. As discussed by \citet{fuselier_etal:14a}, they are either generated inside the instrument due to still ongoing outgassing, or/and a terrestrial-system related foreground. Analyzing the signal from an a priori unknown source and therefore being unable to model it, we first had to decide which data points attribute to the background, and which may be a combination of the background and the signal we seek to analyze. For this, we adopted the simplest assumption: that a flat signal over a contiguous spin angle interval must be background and that the regions near the identifiable signal peak should not be used for background determination. Results of our background determination are in very good agreement with a very sophisticated study by \citet{fuselier_etal:14a}.  
  
\begin{figure}
\epsscale{1.00}
\plotone{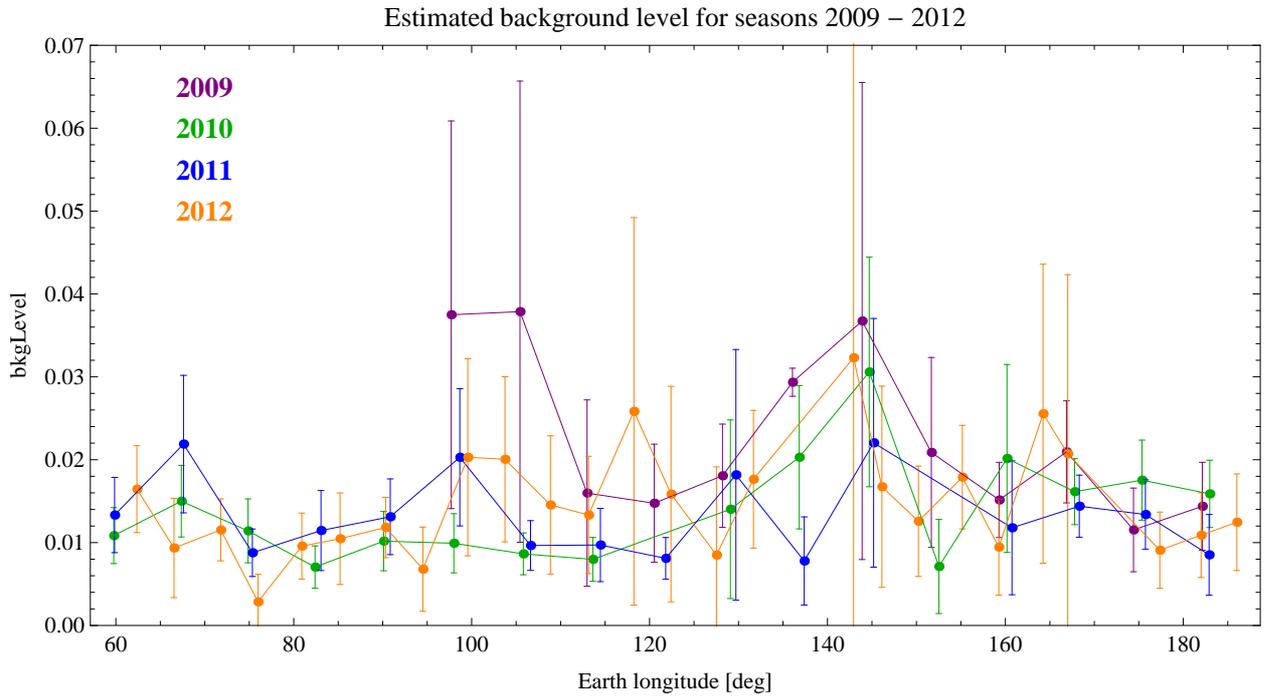}
\caption{Estimated background levels for seasons 2009 through 2012, shown in counts per second per bin as a function of Earth's ecliptic longitude. The units are identical as in Fig.~\ref{fig:fig_dataPlot}.}
\label{fig:bkgLevel}
\end{figure}

Generally, the background was constant in spin angle, except a few orbits. This finding allowed us to adopt as the background level for a given orbit the arithmetic mean of the count rates from the spin angle region away from the signal. For the uncertainty of the background, we took standard deviation of the mean. In a few cases when the background was not flat, we adopted as the background level the minimum value, obtained from a parabolic fit. The background level seems largely repetitive year by year, with a few exceptions. A comparison of background levels for observation seasons 2009 through 2012 is presented in Fig.\ref{fig:bkgLevel}.

The process of data selection was carried out for each orbit separately. We analyzed total counts accumulated during good times on a given orbit. For an orbit $k$, after determining the background level $b_k$ and its standard deviation $\sigma_{b,k}$, we conservatively rejected all data points exceeding $b_k + 2 \sigma_{b,k}$, i.e., at the 95\% significance level. Since the total number of spin angle bins is 60 and the signal typically occupies $\sim 20$ bins, the statistics based on the remaining $\sim 40$ points is not sufficient to eliminate the background at a higher significance level.  

With the background points rejected, we calculated the NIS count rate $c_{ik}$ subtracting the background \textit{b$_k$} from each datum $d_{ik}$ and dividing the result by the length of the good times accumulation interval $t_k$, obtaining
\begin{equation}
\label{rown1}
c_{ik}=(d_{ik}-b_k)/t_k 
\end{equation}
The $c_{ik}$ count rates were subsequently used in model fitting. The errors were calculated assuming that the NIS flux was Poisson-distributed and the background distribution was normal. The uncertainty $\sigma_{c,ik}$ was computed as: 
\begin{equation}
\label{rown2}
\sigma_{c,ik}=\left[ \left(\sqrt{t_k\, c_{ik}}\right)^2+\sigma_{b,k}^2+\sigma_{q,{ik}}^2\right]^{1/2}/t_k
\end{equation}
where $\sigma_{q,{ik}}$ is the uncertainty of the statistical correction $q_{ik}$, conservatively adopted as:
\begin{equation}
\label{rown3}
\sigma_{q,{ik}}=\sqrt{\frac{1}{2}q_{ik}}
\end{equation}

The 267 data points adopted for fitting after background subtraction are shown as dark blue symbols in Fig.~\ref{fig:fig_dataPlot}, while the background level adopted for each orbit is marked with horizontal bar. A consistency check performed after fitting the model showed that our conservative approach to selecting the data was basically correct, i.e., we were able to fit the model to almost all points adopted for analysis, with few outliers in the wings. In many cases, the model even fitted well to points that we left out from the analysis because of suspected background contamination. 

\section{Warm Breeze model and parameter search}
\citet{bzowski_etal:12a} reported that the signal from orbits 61 through 68 corresponds to the NIS He flow and pointed out that an additional signal is visible on earlier orbits. Indeed, as illustrated in Fig.~\ref{fig:fig_peakFlux}, the peaks of the actually observed signal on consecutive orbits 54 through 60 are up to an order of magnitude higher than the peaks of the simulated signal from the NIS He flow. The extra signal is present during the entire duration of IBEX mission until now, as illustrated on the first seven panels of  Fig.~\ref{fig:fig_dataPlot}. In these panels, in addition to the data for the 2010 season, we show normalized count rates for equivalent orbits from seasons 2011 and 2012 (when available). The observations carried out in the 2009 season begun too late during the year to reliably catch the signal corresponding to orbits 54 through 60 from 2010, but the signature of elevated wings of the NIS He distribution is visible during the 2009 season in the orbits when the signal is mostly due to the NIS He population. 

The signal changed only weakly from season to season in the years 2010 through 2012 and thus cannot be due to a transient background. It merges smoothly with the NIS He signal and, as we show below, it can be discerned also in the orbits when the NIS He signal dominates, gradually fading into the background as Earth moves around the Sun. In addition, the time-of-flight spectrometer of IBEX-Lo allows an approximate species determination. As discussed in \citet{mobius_etal:09a} and \citet{mobius_etal:12a}, the signature of the primary NIS He is a contribution of about 20\% C and O in addition to H, knocked out as sputtering products from the conversion surface, in contrast to NIS H, which show as almost 100\% H \citep{saul_etal:12a}. During the early orbits when there is no clear primary NIS He signal, a contribution of about 11\% C and O is found, which is consistent with sputtering by He atoms at noticeably lower energies than the primary NIS He. Thus we adopted a hypothesis that the signal we observe is due to a flow of neutral helium in the heliosphere, which we will refer to as the Warm Breeze. 

The plausibility of the hypothesis that the Warm Breeze is helium was additionally verified by modeling. We assumed that the signal is due to a population of neutral helium that enters the heliosphere from beyond the heliopause. We modeled it as a Maxwellian population that co-exists in the inner heliosphere with the NIS He population without interaction, which is appropriate since the NIS He gas in the heliosphere is collisionless (the mean free path is larger than the size of the heliosphere), so adding an extra tenuous collisionless population does not change this property. In other words, we adopted the following model for the distribution function $f_{He}\left(\vec{v}\right)$ of neutral helium in front of the heliopause:
\begin{equation}
\label{rown4}
f_{He}\left(\vec{v}\right) = n_{NISHe}\left(f_{Maxw}\left(\vec{v},\vec{v}_{NISHe},T_{NISHe}\right)+\xi_{WB}f_{Maxw}\left(\vec{v},\vec{v}_{WB},T_{WB}\right) \right)
\end{equation}
where $f_{Maxw}$ is the Maxwellian distribution function of a monoatomic gas and $\vec{v}$ the velocity vector of individual atom. The formula for $f_{Maxw}$ is the following:
\begin{equation}
\label{rown5}
f_{Maxw}\left(\vec{v},\vec{v}_{B},u_{T}\right) = \left(\frac{1}{\pi u_T^2}\right)^{\frac{3}{2}}\exp\left[-\left(\frac{\vec{v}-\vec{v}_B}{{u}_T}\right)^2 \right]
\end{equation}
The temperature is $T$, corresponding to the thermal speed ${u}_T=\left(2 {k~ T/m}\right)^{1/2}$, with $m$ atomic mass and $k$ the Boltzmann constant. The gas is moving at a vector velocity $\vec{v}_B$. The subscripts NISHe and WB refer to the NIS~He and Warm Breeze components, respectively. The definition \ref{rown4} describes two homogenous populations of neutral gas with different temperatures and different bulk velocity vectors, i.e., flowing with different speeds at an angle to each other. The primary population is the well known NIS He gas, with density $n_{NISHe}=0.015$~cm$^{-3}$ \citep{witte:04, gloeckler_etal:04a, mobius_etal:04a} and temperature, speed, and direction of the bulk flow precisely as found by \citet{bzowski_etal:12a} to minimize $\chi^2$ for the 2010 observation season of NIS He: $v_{NISHe}=22.756$~km/s, $\lambda_{NISHe}=259.2^{\circ}$, $\phi_{NISHe}=5.12^{\circ}$, $T_{NISHe}=6165$~K. These values slightly differ from the values reported by \citet{mobius_etal:12a} and the consensus values worked out by \citet{mccomas_etal:12b} based on the aforementioned papers, which are $v_{NISHe}=23.2\pm 0.3$~km~s$^{-1}$, ${T}_{NISHe}=6300 \pm~390$~K, $\lambda_{NISHe} = 259^{\circ} \pm 0.5^{\circ}$, $\phi_{NISHe} = 5\degr \pm 0.2\degr$. We adopted them because the prerequisite for the parameter search was that the new model fit to the data at least as good as the single-population model fitted to the data subset used by \citet{bzowski_etal:12a} and we knew from the analysis that these parameters correspond precisely to the minimum of reduced $\chi^2$. An illustration of the behavior of the distribution function of the primary population inside the heliosphere is available in \citet{mueller_etal:13a}.

The second term in Eq.~(\ref{rown4}) corresponds to the new Warm Breeze population, for which we search the parameters. The Warm Breeze parameters sought are speed $v_{WB}$, inflow direction ecliptic longitude $\lambda_{WB}$ and latitude $\phi_{WB}$, and temperature $T_{WB}$. Also a free parameter is the abundance of the additional population relative to the NIS He gas, which we denote $\xi_{WB}$. The absolute density of the additional population is determined as a product of $\xi_{WB}$ and the density of the primary population $n_{NISHe}$. Thus we have five parameters to fit and a model defined as a sum of two Maxwellian functions (Equations \ref{rown4} and \ref{rown5}), one of them with known parameters, the other one with parameters to be found. We collectively denote the searched parameters as the vector $\vec{\theta}_{WB} = \left(v_{WB}, \lambda_{WB}, \phi_{WB}, T_{WB}, \xi_{WB}\right)$. The distribution function of this population at IBEX will behave qualitatively similar to the primary population, as illustrated by \citet{mueller_etal:13a}, but details such as the locations of the direct and indirect peaks, the ratios of their heights, and their widths will be different. Because of its observation geometry, IBEX is only able to sample the direct-population peaks from both the NIS primary and the Warm Breeze populations. 

The Warm Breeze parameter search is done using almost identical method as applied by \citet{bzowski_etal:12a}. We calculate the total flux expected from the known NIS He and the new Warm Breeze populations separately for each of the orbits. We use the time-variable ionization rate obtained by \citet{bzowski_etal:13b}, which is essentially identical with the rate used by \citet{bzowski_etal:12a}, including photoionization and charge exchange with solar wind alphas and protons, but with electron impact ionization switched off. We made this choice because the Breeze signal is visible in the He cone region pass well inside 1 AU and over the poles. The current version of the electron ionization model was designed for outside 1 AU, and thus simulating the electron ionization outside the applicability range of the present model might in fact reduce the accuracy of the results. A homogeneous model of electron ionization inside 1 AU and for high heliolatitudes for use in future Warm Breeze studies is under development. When calculating the flux at IBEX, we take into account the actual IBEX-Lo collimator transmission function, the actual velocity of IBEX relative to Earth (varying during an orbit) and Earth relative to the Sun, and the lower energy limit for sputtering by He atoms. We adopt the spin axis of the spacecraft, which determines the field of view on each orbit, as determined by the IBEX Science Operations Center \citep{hlond_etal:12a}. 

The model helium flux at IBEX $F_{He}(k, \psi_i,\vec{\theta}_{WB})$ on orbit $k$ for a spin angle $\psi_i$ is calculated as
\begin{equation}
\label{rown6}
F_{He}(k,\psi_i,\vec{\theta}_{WB})=F_{NISHe}(k,\psi_i)+F_{WB}(k,\psi_i,\vec{\theta}_{WB})
\end{equation}
where $F_{NISHe}(k,\psi_i)$ is the known and fixed contribution from the NIS He population, and ${F}_{WB}(k,\psi_i,\vec{\theta}_{WB})$ is the contribution from the Warm Breeze. The flux $F_{He}$ per unit time and surface area is computed from the distribution function (\ref{rown5}) in the IBEX inertial frame using the standard definition: $F_{He}=\int \int_{u_{lim}}^{\infty} {u}^3\,f_{He}(\vec{v})\, C(\Omega) \mathrm{d}u \,\mathrm{d}\Omega$, where $\vec{u}$ is the velocity vector of a He atom in the IBEX-inertial frame, related to the velocity in the solar inertial frame $\vec{v}$ by $\vec{u} = \vec{v} - \vec{v}_{IBEX}$, where $\vec{v}_{IBEX}$ is the velocity of IBEX relative to the Sun, $u=|\vec{u}|$, and $\Omega$ is the  direction from which the atom is coming to the detector in the IBEX frame. $C(\Omega)$ is the collimator transmission function and $u_{lim}$ the lower detector sensitivity threshold, adopted as 43~km~s$^{-1}$, i.e., the lower boundary for energy step 3. This is because the impacting He atom must have an energy sufficient to sputter H and the lion share of the produced H ions will fall below 20 eV, the lower limit of energy step 2. In addition, at this speed the kinetic energy of He with 38 eV is barely above the energy threshold for sputtering off C, with the remaining sputter efficiency lower by a factor of $\sim 50$ compared with He at the energy (130 eV) of the primary bulk flow speed \citep{yamamura_tawara:96a}, and the contribution to the observed distribution can be neglected. While the exact boundary for sputtering is not precisely known for IBEX-Lo, it does not seem to weigh heavily on the Warm Breeze parameters found from the analysis: the fittting we did assuming the integration boundary 0 brought the Warm Breeze inflow parameters inside our present uncertainty limits.

The upper integration boundary must be formally put to infinity, but in the numerical calculations it was finite and calculated so that less than 0.001 of the total population in the source region is excluded. We verified that increasing this limit did not significantly change the calculated flux. Other details of the flux calculation and the averaging over the collimator transmission function are presented by \citet{bzowski_etal:12a}. 

For convenience, we scale the data and the model so that their respective maxima for Orbit 64 are equal to 1. The maximum of the observed count rate occurs on orbit 64 for spin angle $\psi_0 = 266.5^{\circ}$. The normalization factor for the data is denoted as $c_0$ and for the model $F_0 = {F_{NISHe}(k = 64,\psi_0)+F_{WB}(k=64,\psi_0,\vec{\theta}_{WB})}$. The normalized count rates $c_{i k,norm}$ for comparison with the simulations are calculated as
\begin{equation}
c_{i k,norm}=\frac{c_{i k}}{c_0}\,\mathrm{for\,all\,} i, k
\label{rown7}
\end{equation}
The normalization condition for the model is calculated individually for each parameter set $\vec{\theta}_{WB}$. It must guarantee that the model count rate for spin angle $\psi_0$ on orbit 64 be equal to 1. Thus the normalized flux $F_{He,norm}(k,\psi_i ,\vec{\theta}_{WB})$ is calculated as:
\begin{equation}
F_{He,norm}(k,\psi_i ,\vec{\theta}_{WB})=(F_{NISHe}(k,\psi_i)+F_{WB}(k,\psi_i,\vec{\theta}_{WB}))/F_0
\label{rown8}
\end{equation}
The normalization factor for the model, equal to the term in the denominator in Eq. (\ref{rown8}), changes depending on the parameter values. 

The approach to the parameter search is based on the maximum likelihood method. It involves looking for a parameter set $\vec{\theta}_{WB}$ for which the merit function $\chi^2$ attains minimum value. The merit function is defined as follows:
\begin{equation}
\label{rown9}
\chi^2=\frac{1}{N-M} \sum_{k=1}^{N_{orb}} \sum_{i=1}^{N_{k}} \left( \frac {c_{ik,norm} - F_{He,norm}(k,\psi_i,\vec{\theta}_{WB})}  {\sigma_{ik}   } \right)^2
\end{equation}
where $M =5$ is the number of free parameters, $N_k$ is the number of data points on the $k$-th orbit, $N_{orb}$ is the number of orbits, and $N$ is the total number of data points. Obviously, $N = \sum_{k=1}^{N_{orb}} N_k$. In the absence of analytical form of the model, this minimization was carried out numerically.

\section{Results}

The task of numerical minimization of Eq.(\ref{rown9}) was much more complex than the search for the optimum parameter set for the primary NIS He population. First, the number of points in the present task was larger by almost a factor of 6 (267 vs 46). Second, the calculation time for a single point of the Warm Breeze population was significantly longer than for the NIS He population because the lower inflow velocity and higher temperature required a wider coverage of the IBEX-Lo visibility strip. This is because the angular size of the Warm Breeze beam on the sky is larger. Finally, the slow speed of Warm Breeze atoms implies that they need more time to reach IBEX from the threshold of the heliosphere, which substantially extends the time the atoms must be tracked in the calculation and thus the wall time of the simulations. Also the ionization rate had to be calculated very accurately because the ionization losses for the Breeze atoms are large and the modeled flux depends on the details of the adopted ionization rate more sensitively than the primary NIS He flux. Therefore the minimization process took long and it was not practical to map the 5D parameter space in as much detail as \citet{bzowski_etal:12a} did for the primary NIS He flux (see their Fig. 22). 

\begin{figure}
\epsscale{0.8}
\plotone{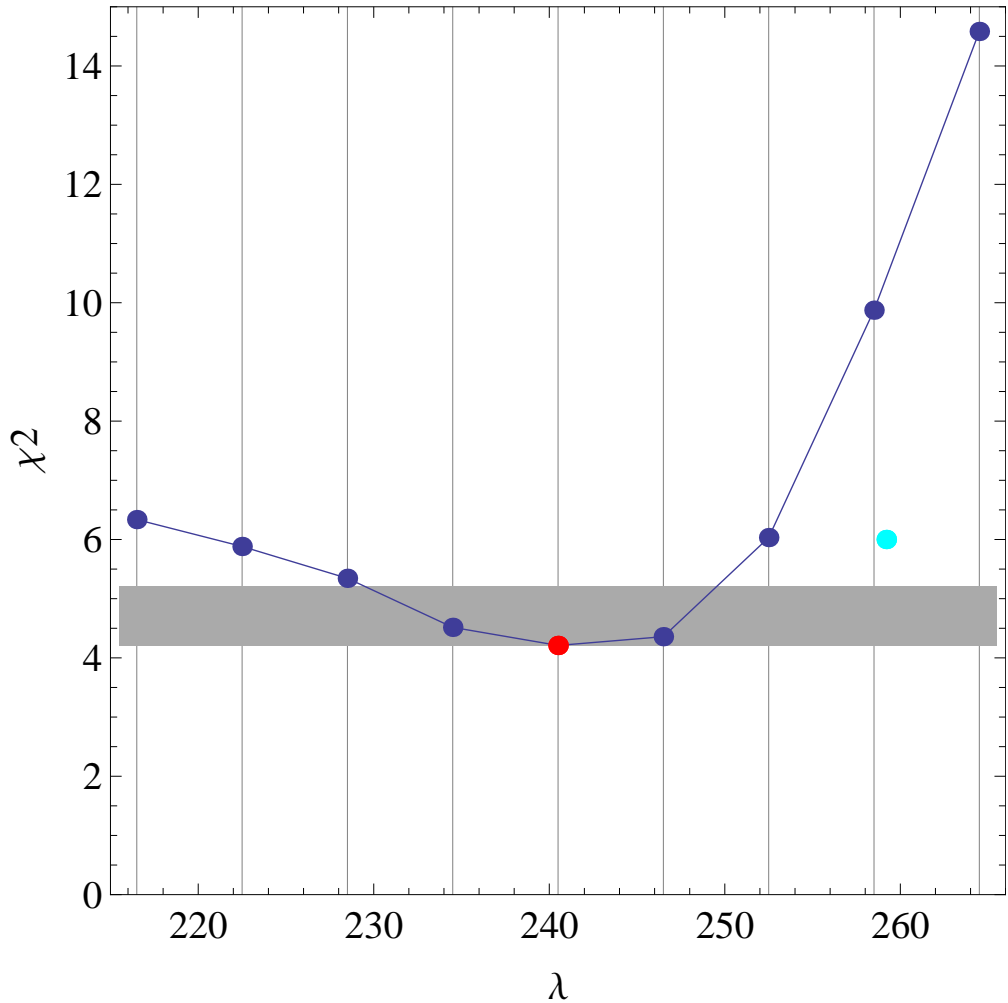}
\caption{Optimized values of the merit function $\chi^2$ for various values of the Warm Breeze inflow direction longitude $\lambda$. The optimum point is marked in red, the cyan point marks the position of the optimum solution found by \citet{bzowski_etal:12a} for the NIS He gas alone, based on the data shown as the red points in Fig.~\ref{fig:figLog_bigComparison_sec_o5459}. The gray bar marks the range of longitudes for which the solutions are considered acceptable: the lower boundary is at the level of the $\chi^2$ minimum, the upper one is on the level of $\chi^2_{min} +1$.}\label{fig:fig2_chi2_lamb}
\end{figure}

The parameter search was organized by ecliptic longitude of the Warm Breeze inflow direction. The optimization was carried out in a 4D parameter space for a number of longitudes, with new longitudes added during the process based on intermediate results. The values of the optimized $\chi^2(\lambda)$ are shown in Fig.~\ref{fig:fig2_chi2_lamb}. The best fitting inflow parameters for the Warm Breeze are temperature 15068~K, speed 11.344~km/s, longitude 240.5$^{\circ}$ and latitude 11.9$^{\circ}$. The density of the Warm Breeze at 150~AU from the Sun relative to the unperturbed NIS He gas in the LIC is 7\%. These values are reported as those that minimize the reduced $\chi^2$; the physical interpretation will be discussed further on in the paper. 

The values of the temperature and velocity parameters seem to be correlated to each other in a similar way as the parameters of NIS He, as found by \citet{bzowski_etal:12a, mobius_etal:12a, lee_etal:12a}. The quality of the solution seems better than the quality of the solution found by \citet{bzowski_etal:12a} for the NIS He population alone, as can be appreciated by comparing the present value of reduced $\chi_{min}^2 \simeq 4$ with the reduced $\chi^2 \simeq 6.5$ from \citet{bzowski_etal:12a} (cf Fig.~\ref{fig:fig2_chi2_lamb}).

\begin{figure}
\epsscale{1.0}
\plotone{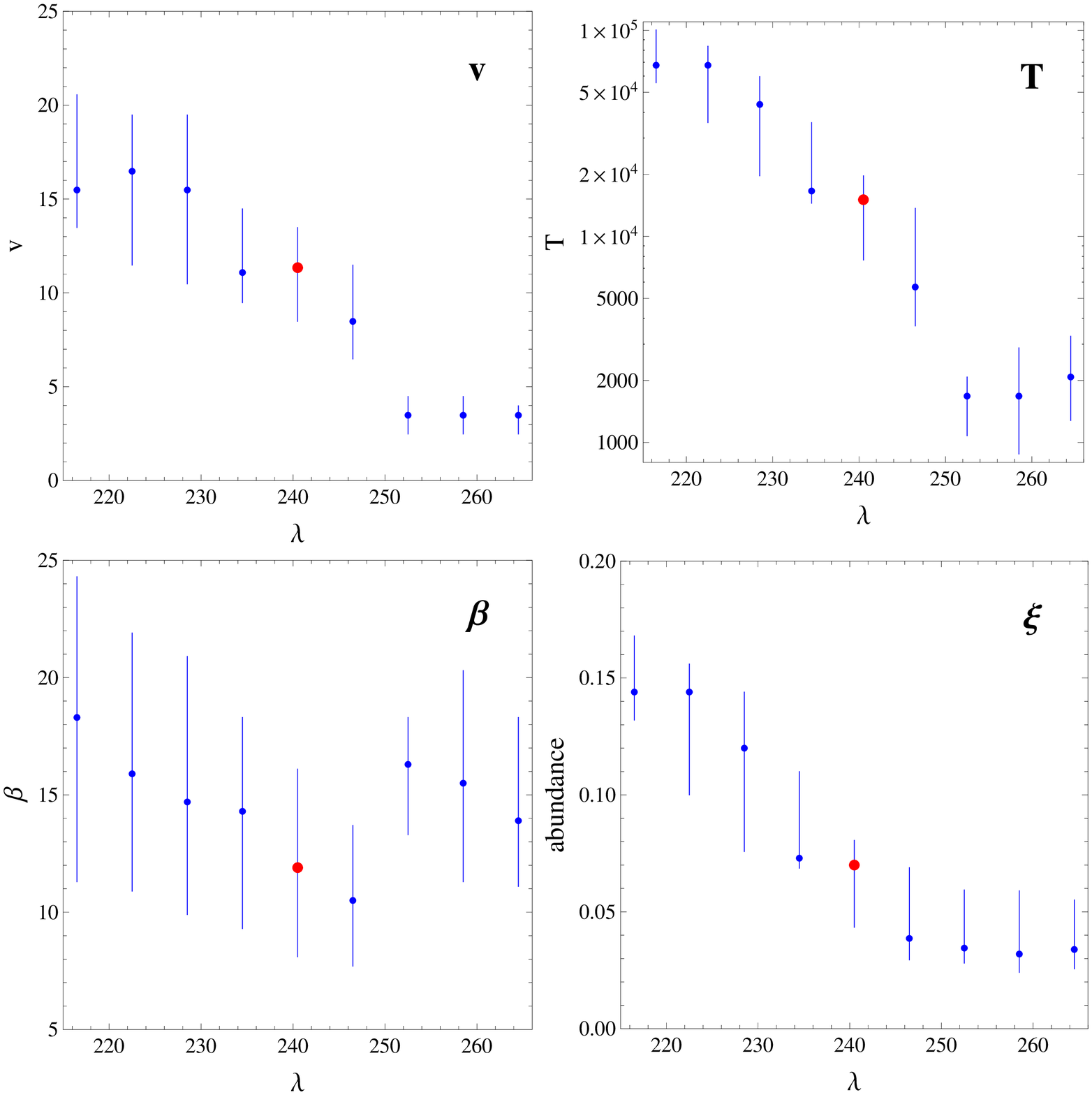}
\caption{Approximate areas in the parameter space where the $\chi^2$ values are within 1 of the minimum value for respective ecliptic longitudes $\lambda_i$, i.e., the acceptable regions for the values of the inflow parameters and density of the Warm Breeze, shown in 2D cuts as a function of ecliptic longitude of the inflow direction. The absolute minimum is marked with the red dot. The blue dots mark the minima in $\chi^2$ calculated for the other assumed $\lambda_i$ values. The vertical bars mark the region within 1 of $\chi^2_{min}(\lambda_i)$.  The parameters shown are inflow velocity (upper left panel), temperature (upper right), inflow latitude (lower left), and density relative to NIS He density (lower right). Note the vertical scale for temperature is logarithmic.} \label{fig:fig_chi2_grid}
\end{figure}

The uncertainties in the found parameters were assessed in the standard way, i.e., by adopting the ranges corresponding to $\chi^2$ values within $\chi^2_{min} + 1$. This level is indicated in Fig.~\ref{fig:fig2_chi2_lamb}. The geometric locations of these parameter values in 2D cuts through the parameter space are shown in Fig.~\ref{fig:fig_chi2_grid}. The acceptable range for longitude is from $\sim 228\degr$ to $\sim 250\degr$ and from $9\degr$ to $15\degr$ for latitude, with the inflow speed varying accordingly from $\sim 15$~km/s down to $\sim 7$~km/s. For this range, temperatures vary from $\sim 30~000$~K down to $\sim 5000$~K, and densities from $\sim 12$\% to $\sim 5$\% of the density of NIS He. 

\begin{figure}
\epsscale{1.0}
\plotone{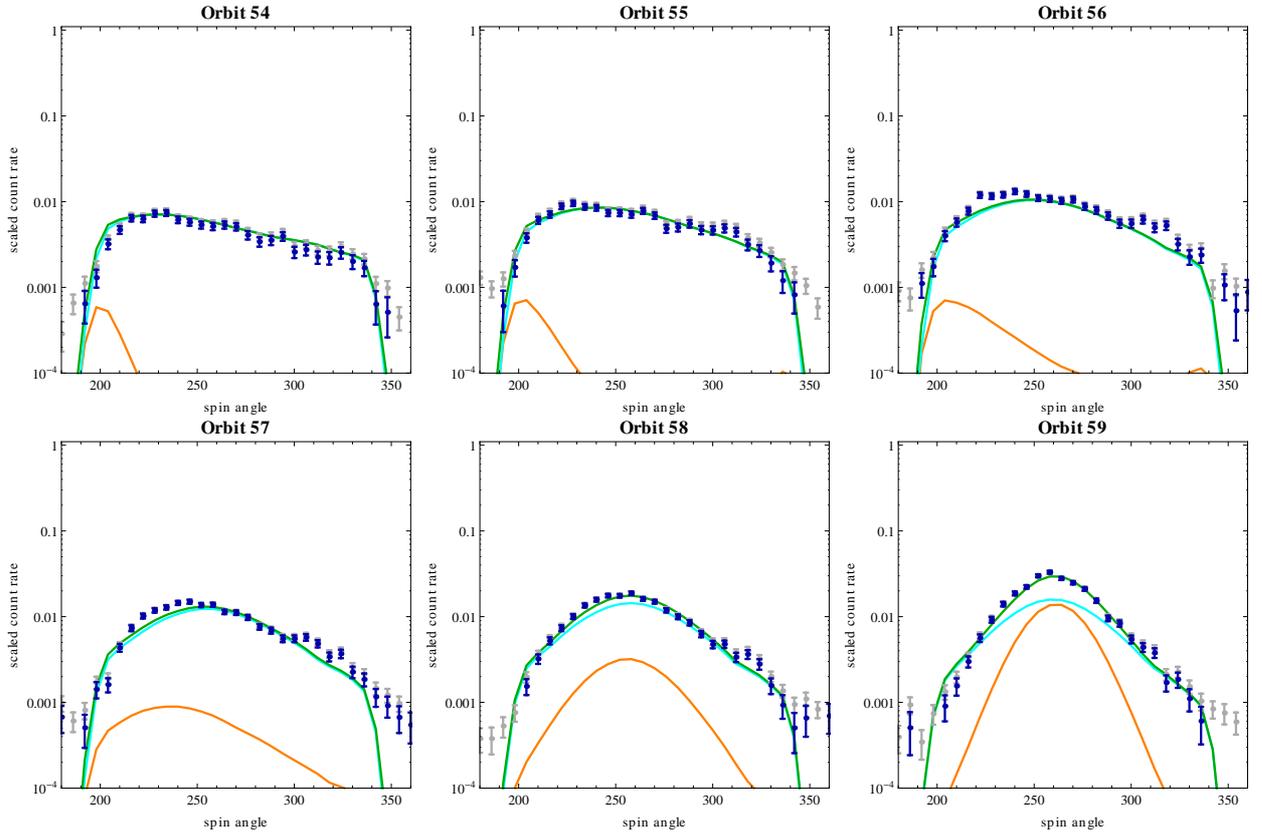}
\caption{Data used for fitting (dark blue symbols with error bars), compared with the best fitting model of neutral helium flux observed by IBEX-Lo (green line), being a sum of fluxes from the NIS He gas (orange) and Warm Breeze populations (cyan). The data and the model are scaled to their respective peak values for orbit 64. Gray symbols represent raw data without background subtracted.}.\label{fig:figLog_bigComparison_sec_o5459}
\end{figure}

The low $\chi^2$ region in the parameter space is much more complex in shape than in the case of the NIS He population examined by \citet{bzowski_etal:12a}. This may be because the homogeneous Maxwellian flow model is not fully adequate to describe the Warm Breeze. Still, the best-fit solution reproduces the data very well indeed, as can be appreciated in Fig.~\ref{fig:figLog_bigComparison_sec_o5459}. The superposition of the NIS He population from \citet{bzowski_etal:12a} and the Warm Breeze discovered now explains the entire signal above the background in orbits 54 to 68. It explains as well the points corresponding almost solely to the NIS He population, as the points that \citet{bzowski_etal:12a} could not interpret in their analysis. Even some of the data points rejected from the analysis as suspect of background contamination seem to fit well to the model (e.g., orbit 63, 64, 66). The simulated total signal matches reasonably well the observations from orbit 69, which were not used in the fitting because they show a non-flat background and low signal to noise ratio, and because a small contribution from NIS H is expected in this region of the Earth orbit in energy step 2 (most of the signal from NIS H is in energy step 1; \citet{saul_etal:12a, schwadron_etal:11a}). The increasingly non-flat background and NIS H contribution are the reasons why we restricted the data used for fitting the Warm Breeze to orbits earlier than 69.

Based on the analysis presented above we conclude that the signal observed by IBEX-Lo on orbits 54 through 69 and corresponding orbits from the other observation seasons is consistent with two populations of neutral helium flowing into the heliosphere: neutral interstellar helium from the Local Interstellar Cloud, and an additional population that we call the Warm Breeze, whose origin is not fully understood yet.

\begin{figure}
\epsscale{1.0}
\plotone{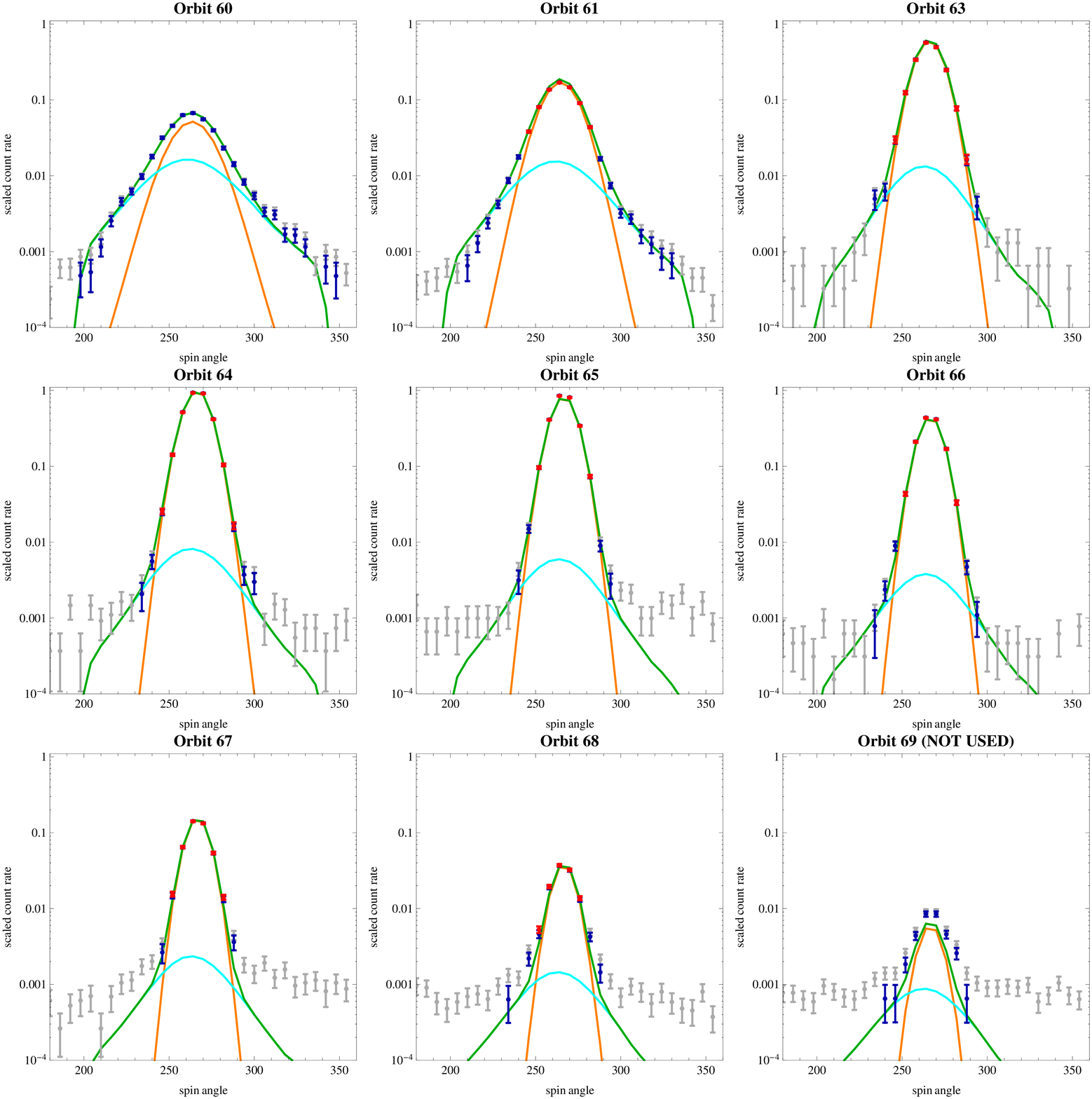}
\caption*{Fig. 6. continued \textemdash: Red points represent the data used by \citet{bzowski_etal:12a} in the analysis of NIS He flow. The data for Orbit 69 were not used in the fitting of the Warm Breeze in this work and by \citet{bzowski_etal:12a} in the fitting of  NIS He.}.\label{fig:figLog_bigComparison_sec_o6069}
\end{figure}

\section{Discussion}

\subsection{Can the observed signal be explained without the Warm Breeze?}

The neutral He signal observed by IBEX-Lo was first interpreted as due to the primary NIS He population, but unexplained remained the elevated wings in ecliptic latitude and the additional peak seen at ecliptic longitudes preceding the longitude interval occupied by the NIS He signal. 

\citet{gruntman:13a} suggested that the elevated wings may be due to a broadening of the NIS He beam inside the heliosphere due to elastic collisions between some of the inflowing He atoms with solar wind protons and alpha particles. He estimated that a departure from the Gaussian shape of the signal from the Maxwellian NIS He gas as a function of ecliptic latitude (equivalent to spin angle) on a given orbit due to the collisional broadening should start approximately three orders of magnitude below the peak value. The departures of the signal from the Gaussian shape that we see in the data start at a somewhat higher level, but this alone could possibly be explained by a collision rate higher than \citet{gruntman:13a} allowed for. However, the collisions of NIS He atoms with solar wind ions cannot explain the peak in the signal that we observe at the early orbits each season. Since the Warm Breeze is able to explain both the elevated wings and the peaks of early obits, we consider the Warm Breeze existence as a more likely explanation for the entirety of the observed signal than the elastic collisions hypothesis. The Warm Breeze existence does not rule out a collisional broadening of the signal, but the magnitude of the broadening effect must certainly be lower than the level of the Warm Breeze signal. 

The data we see are also not consistent with a single NIS He population with a non-Maxwellian  distribution function in front of the heliopause, e.g., a kappa function. Whatever the form of the distribution function may be, the prerequisite is that the core of the function do not significantly depart from the Maxwellian function because the NIS He signal interpreted by \citet{bzowski_etal:12b} using the Maxwellian function matches the high core of the observations very well. To check if the signal we are investigating here can be explained by a single-population kappa function, we adopted the following definition of the kappa function (for unit density) and we simulated the expected signal for orbits 54 through 68:
\begin{equation}
f_{\kappa}(\vec{v}) = \left( \frac{1}{\pi u_T^2}\right)^{\frac{3}{2}} 
\left[ 1+ \frac{\left(\vec{v}-\vec{v}_{B}\right)^2}{ \left(\kappa-\frac{3}{2}\right)u_{\kappa}^2}      \right]^{-1-\kappa}
\label{rown10}
\end{equation}
where $u_{\kappa}^2 = u_T^2 \frac{\kappa+1}{\kappa-\frac{3}{2}}$. Note that the normalization term we adopted differs from the normalization usually taken for the kappa function \citep{livadiotis_mccomas:13a}. We did so because we wanted the peaks of the Maxwellian and kappa distributions to agree with each other. A change in the normalization does not change the shape of the function, only the absolute height of the peak, which is not important here because of the normalization of the data and simulations we used in the fitting.

\begin{figure}
\epsscale{1.0}
\plotone{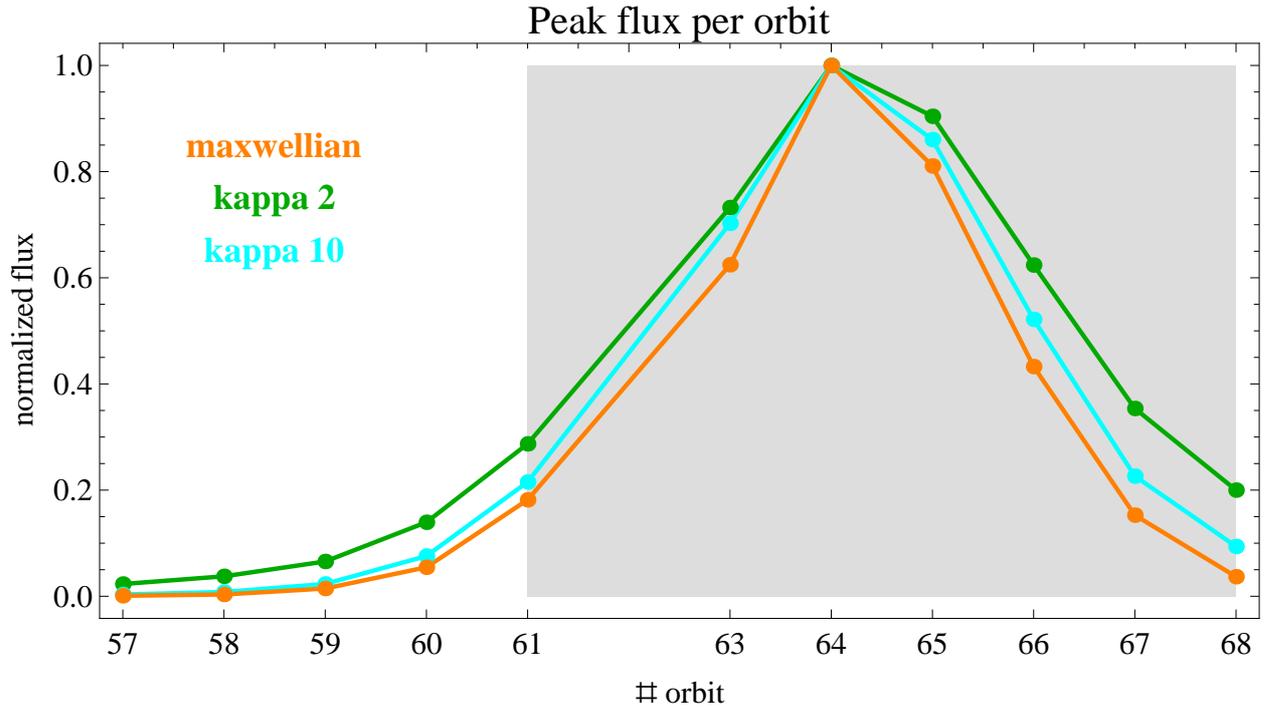}
\caption{Comparison of simulated peak heights for the orbits used in the analysis for a Maxwellian distribution function with the parameters corresponding to the parameters of the NIS He gas found by \citet{bzowski_etal:12a} (orange) with simulated peak heights for identical parameters, but a $\kappa$  distribution function given by Eq. (10), with $\kappa = 10$ (cyan) and 2 (green). The shaded area corresponds to the orbits analyzed by \citet{bzowski_etal:12a}, where the signal from a $\kappa$ function must agree with the Maxwellian signal.}\label{fig:figLin_priKappa_peakSeries}
\end{figure}

A comparison of the simulated peak heights for orbits 54 through 68 for the Maxwellian distribution function and the parameters obtained by \citet{bzowski_etal:12a} with the peak heights of the flux simulated for a kappa distribution function with the flow parameters identical as for the Maxwellian function and kappa values decreasing from the high $\kappa$~=~10 (i.e., almost Maxwellian) down to $\kappa$~=~2 (strongly non-Maxwellian) is shown in Fig.~\ref{fig:figLin_priKappa_peakSeries}. For the Warm Breeze signal to be explained by a single kappa-like distribution function of the NIS He gas in front of the heliopause, first the peak heights for the orbits analyzed by \citet{bzowski_etal:12a} must agree with the peaks calculated using the Maxwellian function. If this test is passed, then another will be the agreement of the signal as a function of spin angle. 

Inspection of Fig.~\ref{fig:figLin_priKappa_peakSeries} shows that even for the moderate $\kappa$~=~10 case, the simulated signal does not match the Maxwellian case in the region marked in the figure as corresponding to the orbits used in the analysis by \citet{bzowski_etal:12b}. Thus, the test for kappa function fails at the most elementary level, i.e., the peak heights, which should be the least sensitive to the non-maxwellian nature of the distribution function. Reducing the $\kappa$ value increases the disagreement. Hence we conclude that our signal cannot be explained by a single Maxwellian or kappa-like population in the LIC. 

Therefore an extra population of neutral He gas is needed to explain the data. Elevated wings in the signal visible on the orbits where NIS He is dominant may be explained by an additional population, Maxwellian or perhaps kappa-like, coming from the same direction as the NIS He gas, but the signal visible on the early orbits each year cannot. Therefore the additional population must come from a direction different from the direction of NIS gas.

\subsection{Is the Warm Breeze signal consistent with helium inflow?}

The Warm Breeze signal is detected as H$^-$ ions in energy steps 1, 2, and 3 of IBEX-Lo. The simulations we performed suggest that the signal is consistent with these H$^-$ ions being sputtered by neutral He atoms impacting at the IBEX-Lo conversion surface. The signal we see is from recoil sputtering. In this mechanism, the top layer (i.e., water in the case of the IBEX-Lo conversion surface) is removed by particles impacting at shallow impact angles and the sputtering products must be dissociated to produce a populations of H$^-$ ions. This process involves a low energy threshold, especially for negative ions \citep{taglauer:90a}.

When looking for the Warm Breeze inflow parameters, we assumed in the simulations that the signal is from a sputtered component, so, based on a sputtering threshold studies by \citet{yamamura_tawara:96a, taglauer:90a}, the particles impacting at the conversion surface of IBEX-Lo must have energies exceeding $\sim$10 -- 30~eV. On the other hand, as indicated by \citet{mobius_etal:12a}, once the impactor has an energy sufficient to sputter, the energy distribution of the sputtering products depends very weakly on impactor’s energy. 

\begin{figure}
\epsscale{1.0}
\plotone{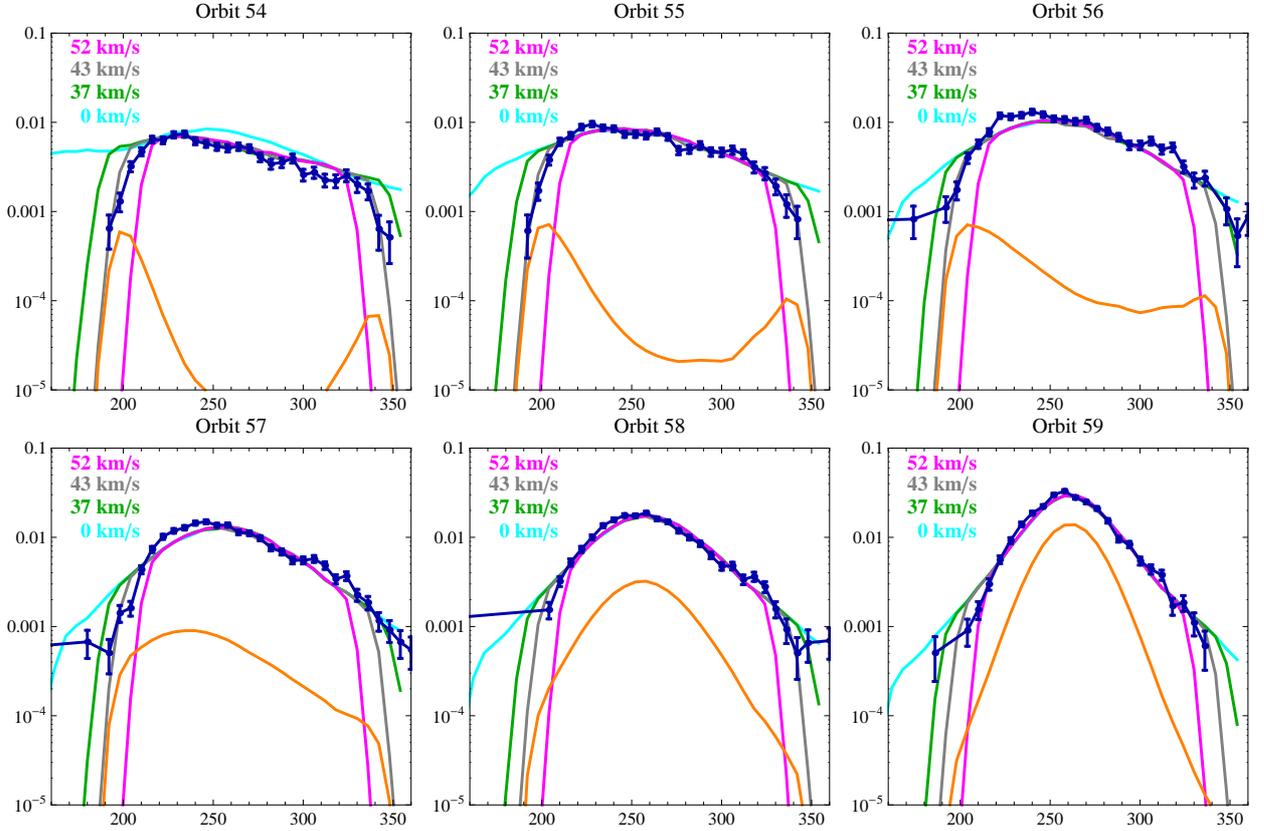}
\caption{Data vs simulations (NIS He + Warm Breeze) differing by the lower limit of the integration boundary over speed. The simulations were carried out for the full velocity range from 0 relative speed in the IBEX inertial frame (cyan), and for the lower integration limits of 37~km/s (green), 43~km/s (gray), and 52~km/s (magenta). Dark blue denotes the data. The limit for the case shown in gray was actually used in the fitting. Orange is the simulated flux for the NIS He component. The horizontal axis in all panels is the spin angle, the vertical scale is the flux normalized similarly as in Fig.~\ref{fig:figLog_bigComparison_sec_o5459} }\label{fig:figLog_energyGrid_secondary}
\end{figure}

The simulated flux is a line of sight integral over velocity in the IBEX inertial frame and the integration boundaries span from the lower sputtering limit equivalent to 43~km/s to the upper limit obtained as velocity in the LIC differing from the inflow velocity by no more than 4.5 thermal speeds, gravitationally accelerated from the source region at 150~AU down to 1~AU. The agreement of the shape of the simulated flux as a function of spin angle with the observations was excellent for all orbits. To verify that the sputtering cutoff really exists, we performed an additional series of calculations, varying the lower integration boundary. The results are shown in Fig.~\ref{fig:figLog_energyGrid_secondary}. It is evident that the simulation performed with the lower integration boundary corresponding to the sputtering limit fits best to the data. Too low or too high an integration boundary results in a misfit at the wings, and the dropoff of the signal with spin angle depends sensitively on the integration boundary. The simulation with the integration boundary 0 does not fit at the wings at all, by orders of magnitude. 

The non-zero integration boundary was introduced to the simulation scheme for the purpose of this study. \citet{bzowski_etal:12a} integrated starting from 0 relative speed. We have verified, however, that for the inflow velocities of 22.8 to 26.3~km/s and temperatures on the order of 6000 -- 7000~K, as in the NIS He flow solutions obtained by \citet{bzowski_etal:12a} and \citet{witte:04}, respectively, the percentage of He atoms impacting IBEX with relative velocities below the sputtering limit is so low that it does not affect the result of the simulation, so the conclusions drawn by \citet{bzowski_etal:12a} remain valid. 

We conclude from this study that the signal observed is consistent with H$^-$ ions sputtered from the conversion surface by neutral He atoms. Thus, we believe the Warm Breeze most likely is helium.

\subsection{Can the Warm Breeze signal be explained by the heliospheric cone of neutral interstellar helium?}

\begin{figure}
\epsscale{1.0}
\plotone{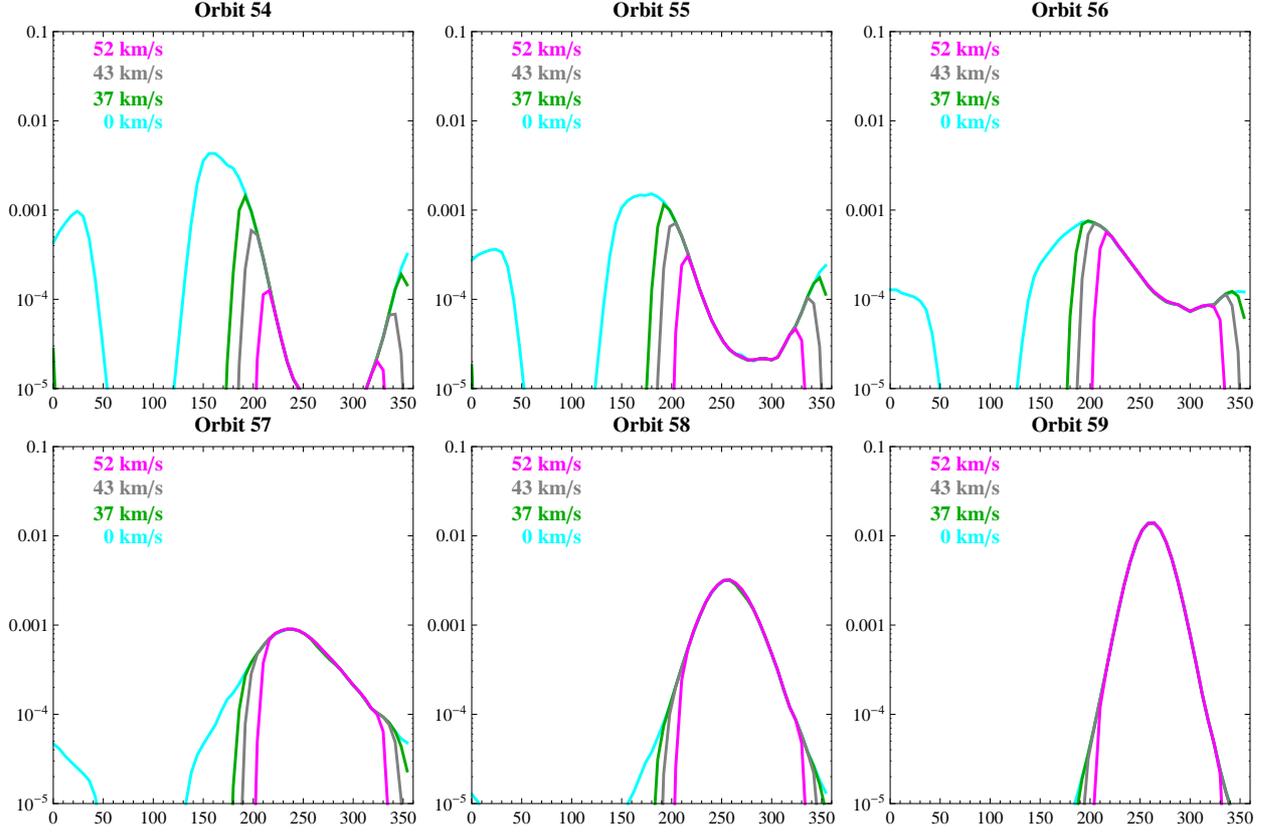}
\caption{Expected signal from the NIS He population for orbits 54 through 59, i.e., from the crossing of the NIS He cone region by IBEX for various lower boundaries of integration of the flux in the IBEX-inertial frame over speed. The signal is scaled to the peak flux on orbit 64, identically as in Fig.~\ref{fig:figLog_energyGrid_secondary}, the vertical scales are also identical. Note that the vertical axis is shifted downwards relative to Fig.~\ref{fig:figLog_energyGrid_secondary}. The color coding and axes are  identical as in Fig.~\ref{fig:figLog_energyGrid_secondary}.}\label{fig:figLog_energyGrid_primary}
\end{figure}

Once we have verified that the Warm Breeze is neutral helium, we must check whether it can be explained by NIS He atoms from the cone of increased density of interstellar He atoms that is formed behind the Sun due to the action of solar gravity \citep[e.g.,][]{fahr:71}. It is obvious that the Warm Breeze is observed as the only population seen in the data when Earth passes through the heliospheric helium cone each year, so an immediate question is whether it can be explained by the cone atoms. 

Most of the cone atoms, however, cannot be observed by IBEX-Lo since they cannot enter the collimator because of the observation geometry, and those few that can, are too slow in the spacecraft reference frame to be registered, unless the instrument is much more sensitive to low-energy He atoms than we believe. This qualitative argument is in full agreement with simulation results. The levels of the signal expected from NIS He and actually observed are shown in Figures \ref{fig:fig_peakFlux} and \ref{fig:figLog_bigComparison_sec_o5459}, and details of expected NIS He signal for various integration boundaries, analogous with the parametric study of the flux dependence on the lower energy limit presented in the preceding section, are illustrated in Fig.~\ref{fig:figLog_energyGrid_primary}.

The pre-flight sensitivity calibration of IBEX-Lo seems to be correct, as objectively supported by the lack of detection of NIS He gas during the other potential NIS He gas observation season early Fall each year, when the alignment of the IBEX detectors with the NIS He flow is favorable for viewing. Despite the favorable geometry, since the Earth travels along with the flow of the gas, the energy of atoms relative to IBEX is much below the sputtering limit of $\sim$10 -- 30~eV. The signal level expected from the NIS He cone is 2 orders of magnitude below the observed level and does not fit at all to the observed shape. 

Thus we conclude that the Warm Breeze signal cannot be due to the heliospheric helium cone. 

\subsection{Why has Warm Breeze not been detected earlier?}

\begin{figure}[hb!]
\epsscale{1.0}
\plotone{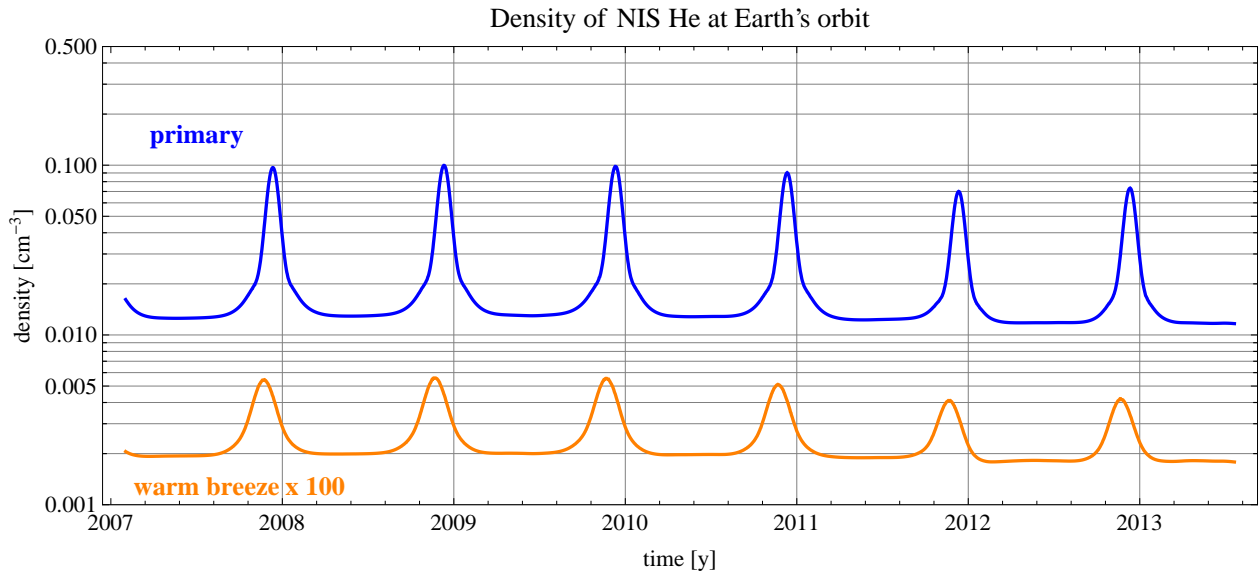}
\caption{Evolution of absolute density of NIS He gas (blue) and the Warm Breeze (orange) at Earth from 2007.1. Note the Warm Breeze line is multiplied by a factor of 100.}.\label{fig:fig_densNISHe}
\end{figure}

None of experimental studies of NIS He and its derivative populations in the heliosphere has discovered the Warm Breeze before IBEX. Answering the question why may help answer the question if the Warm Breeze is a permanent feature or a kind of transient phenomenon. To address this question, we calculated the density evolution of the NIS He gas at Earth from 2007, assuming the inflow parameters as reported by \citet{bzowski_etal:12a}, and of the Warm Breeze, assuming the inflow parameters as found in the present paper. For the Warm Breeze density at the heliopause we adopted the 7\% abundance of the Breeze population relative to NIS He density equal to 0.015~cm$^{-3}$ \citep{gloeckler_etal:04b, witte:04, mobius_etal:04a}, i.e., 0.07 $\times$ 0.015~=~0.001~cm$^{-3}$. The calculations were carried out using the time-dependent Warsaw Test Particle Model for neutral helium gas distribution in the heliosphere \citep{rucinski_etal:03}, with the ionization rate based on actual observations of the ionization factors in the heliosphere from \citet{bzowski_etal:13b}, including the electron impact ionization \citep{rucinski_fahr:89, bzowski_etal:13b}. The results are shown in Fig.~\ref{fig:fig_densNISHe}. 

The density of the Warm Breeze gas at 1~AU from the Sun is $\sim$3 orders of magnitude lower than the density of NIS He, and the ratio of the yearly-averaged densities does not vary significantly with solar activity. The slow inflow speed exposes the Breeze atoms to ionization losses inside the heliosphere much stronger than the NIS gas atoms, which results in a very strong attenuation of the Breeze before it reaches 1~AU. The signals of the EUV backscatter and pickup ion observations are proportional to the local density of the neutral helium gas in the heliosphere, so the respective signals expected from the Breeze are several orders of magnitude lower than the corresponding signals from the NIS He population and thus not detectable. This conclusion holds for the entire region of acceptable parameters for the Warm Breeze, as we verified by modeling the cases from the opposite ends of the acceptable region shown in Fig.~\ref{fig:fig2_chi2_lamb}. 

Another potential opportunity to discover the Warm Breeze could have been through Ulysses/GAS direct-sampling. We have simulated expected signals due to the Warm Breeze for the actual conditions of Ulysses/GAS observations \citep{witte:04} and compared them with simulated signals due to the NIS He flow. We found that while the abundance of the Warm Breeze relative to NIS He at Ulysses was more favorable than at Earth, the signal level was still more than one order of magnitude below the background level. In addition, the maximum of the Warm Breeze beam was most of the times outside the GAS field of view. Therefore it was extremely challenging, if at all possible, to detect the Warm Breeze signal in the Ulysses/GAS data.

Based on these studies we conclude that the Warm Breeze escaped detection up to now because of sensitivity limitations of past pickup ion, EUV backscatter, and direct sampling observations. It could now be discovered owing to the unique capabilities of IBEX-Lo, in particular its high signal to noise ratio as compared to previous instruments. 

\section{Speculation: What is the source of the Warm Breeze?}

In this section we speculate on a possible source for the Warm Breeze, without in-depth elaborating on the hypotheses we put forward. Verification of these hypotheses will be a subject of future studies. 

We propose two groups of hypotheses: (1) the Warm Breeze is related to the heliosphere and is created due to processes operating in the interface region between the heliosphere and the LIC, and (2) the Warm Breeze is due to departures of a portion of the material in the local Galactic environment of the Sun from collisional equilibrium. Within group (1), we consider hypothesis (1.1), that the Warm Breeze is indeed the secondary population of NIS He, created in the outer heliosheath, and alternatively (1.2) that it is created due to elastic scattering of heliospheric ENAs beyond the termination shock on the atoms from NIS He gas. Within group (2), we wonder if (2.1) the Warm Breeze may be a non-thermalized gust from a nearby boundary between the LIC and G cloud or (2.2) due to disturbances or wave trains in the LIC plasma, imparted to neutral helium and traveling distances on the order of $< 10^4$~AU. 

\subsection{Warm Breeze is related to the heliosphere}
\subsubsection{Warm Breeze is the secondary population of NIS He}

\citet{bzowski_etal:12a} hypothesized that the population they discovered, which we now call the Warm Breeze, is likely the secondary population of NIS He. This population is expected to be created in the outer heliosheath as a result of charge exchange between NIS He atoms and charged particles in this region. \citet{bzowski_etal:12a} pointed out that the most efficient source of the secondary population should be charge exchange between the interstellar He atoms and interstellar He$^+$ ions, exceeding by an order of magnitude the intensity of charge exchange between He atoms and protons because of the huge difference in the reaction cross sections. 

To assess the expected distribution of the secondary interstellar helium flow, we employed the global 3D kinetic-MHD Moscow model of the solar wind interaction with the LIC (e.g., \citet{izmodenov_etal:09a}). In these particular calculations, we took into account the resonance charge exchange process of helium  (He$^+$ + He $\rightarrow$ He + He$^+$) in the outer heliosheath. To calculate the distribution of interstellar helium ions, we assumed that their bulk velocity and temperature are equal to the velocity and temperature of  the proton component in the unperturbed LIC, and solved the continuity equation for a two-component plasma and two-component neutral gas, with H and He treated separately.  For neutral interstellar helium, we solved the kinetic equation similar to the equation for interstellar H atoms (see, e.g., Eq.~(4.8) in \citet{izmodenov_baranov:06a}), taking into account the corresponding charge exchange cross section. Further details can be found in Appendix \ref{sec:appendixA}. 

\begin{figure}
\epsscale{0.7}
\plotone{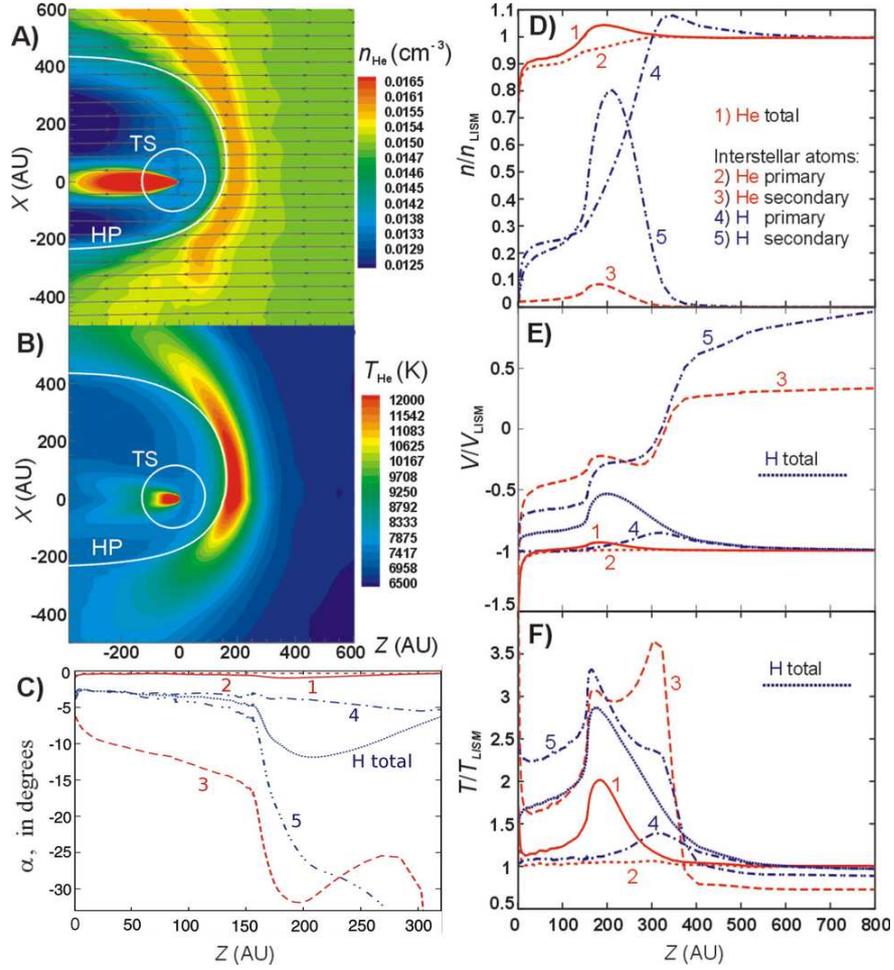}
\caption{Neutral He and H in the heliosphere based on the Moscow Monte Carlo kinetic model of the global heliosphere. Panel A shows the distribution of density of neutral helium in front of and inside the heliosphere in the plane determined by the inflow direction and interstellar magnetic field vector. The gas flows from the right-hand side along the Z axis. Magnetic field is inclined at $20\degr$ to this axis. The termination shock (TS) and heliopause (HP) are presented as white contours. Panel B shows the corresponding distribution of temperature. Panels D through F present profiles of density (D), speed (E) and temperature (F) along the inflow axis for various populations of neutral H and He: primary interstellar H and He, secondary H and He, and the combined primary and secondary populations of H and He. For the legend, see panel D. The quantities are shown as ratios to the respective values in the LIC. Panel C presents the angle between the velocity vectors of the primary and secondary NIS He and H populations along the z-axis on one hand and the direction of inflow on the other hand. The directions of the averaged flows, calculated as the angle between the local average velocity and the direction of the flow in the unperturbed LIC, is also shown.} \label{fig:figure_He2new}
\end{figure}

The interstellar parameters were adopted as follows: the inclination of the interstellar magnetic field direction to the gas inflow direction $\alpha_B(\vec{B}_{LIC},\vec{v}_{LIC}) = 20^{\circ}$, the strength of interstellar magnetic field ${B}_{LIC} = 4.4~\mu$G, the speed of the interstellar gas inflow on the heliosphere $v_{LIC} = 26.4$~km/s, temperature $T_{LIC} = 6530$~K, proton density in the unperturbed interstellar medium $\rho_{LIC} = 0.06$~cm$^{-3}$, neutral hydrogen density $n_{H,LIC} = 0.18$~cm$^{-3}$, and neutral helium density $n_{He,LIC} = 0.015$~cm$^{-3}$. These parameters are essentially identical as those used by \citet{bzowski_etal:08a} to infer the density of interstellar neutral H at the termination shock based on Ulysses observations of pickup ions. The solar wind was assumed spherically symmetric. Note that with this set of parameters a bow shock does not form in front of the heliosphere because of the high strength of interstellar magnetic field assumed. The orientation of the magnetic field vector was chosen to put it into the Hydrogen Deflection Plane \citep{lallement_etal:05a, lallement_etal:10a}. As discussed by \citet{izmodenov_etal:09a}, with the adopted magnitude $B_{LIC}$ and inclination $\alpha_B$ of the interstellar field as well as the other interstellar parameters, the distances to the termination shock obtained in the simulation were in agreement with the actual distances measured by the Voyager spacecraft \citep{stone_etal:05a, stone_etal:08a}. Further discussion of the interstellar parameters can be found in \citet{izmodenov_etal:09a}. The goal was to take a set of reasonable parameters and see how the secondary population of NIS He would flow in the outer heliosheath. Such simulations for helium, to our knowledge, have never been performed before. The resulting densities, speeds, and temperatures of the interstellar and secondary populations of NIS H and He are presented in Fig.~\ref{fig:figure_He2new}.

The behavior of neutral helium in the heliospheric interface is qualitatively similar to the behavior of neutral H, but quantitative differences are obvious. Similarly to H, the secondary He develops a ``wall'' of increased density in front of the heliopause, which is offset upstream by a dozen of AU. Such a wall was theoretically predicted by \citet{baranov_etal:91} for hydrogen and discovered by observations of interstellar absorption lines by \citet{linsky_wood:96}. The bulk speed of the secondary component in this region is on the order of 0.25 to 0.3 of the inflow speed, and the temperature is increased by a factor of 3 to 3.5. The density is approximately 0.07 of the density in the unperturbed LIC. Inside the heliopause, the bulk speed of the secondary component increases to approximately half of the speed in the LIC, and temperature drops to $\sim$1.6 of the temperature in the LIC. The density in this region is only $\sim$0.01 of the density of the unperturbed He.  The direction of flow of the secondary population at the upwind axis relative to the inflow velocity varies within the Helium Wall, but attains $\sim 30\degr$~in the region of maximum density. Inside the heliopause, this offset in direction is reduced to $\sim 15\degr$. An important factor is that all parameters of the secondary population are not homogeneous in the source region in front of the heliopause, featuring gradients both in radial distance from the Sun, and in the direction. 

The speed and temperature of the secondary He component, as well as its density relative to the density in the LIC qualitatively agree with the parameters we obtained for the Warm Breeze, so in this aspect, the modeling we performed favors the hypothesis that the Warm Breeze is in fact the heliospheric secondary population of NIS He. Also the offset of the inflow direction of the Warm Breeze may agree with the angles between the local flow of the secondary population from the model to the upwind direction of interstellar gas. But what can immediately be noticed by inspection of Fig.~\ref{fig:figure_He2new} is the spatial variations of the secondary population parameters. 

\begin{figure}
\epsscale{1.0}
\plotone{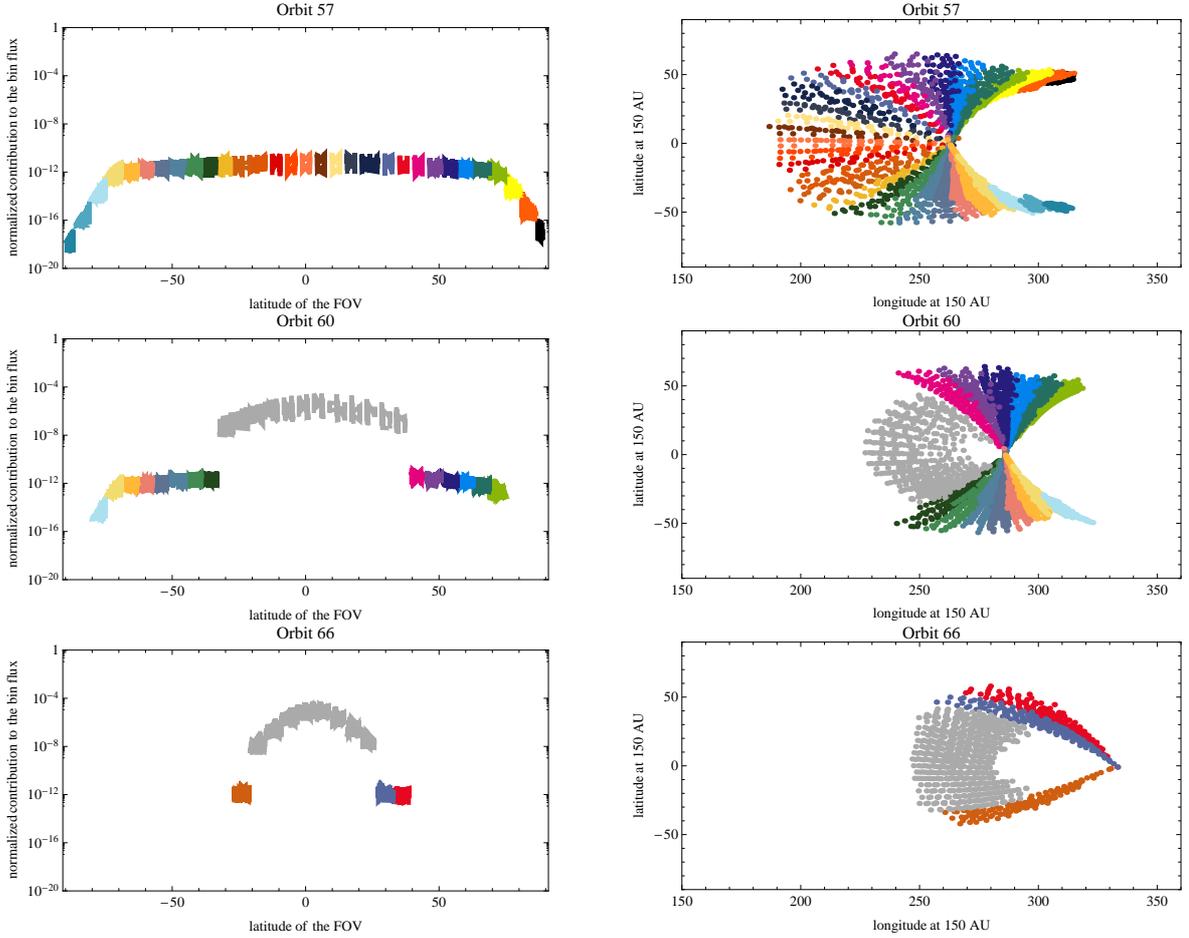}
\caption{Illustration of the correspondence between the directions of IBEX-Lo pixels and geometric locations of the crossings of a 150~AU sphere centered on the Sun by neutral He atoms contributing to the signal. Each Warm Breeze bin is marked with a different color. In the left-hand column, we present contributions to the flux for a given spin angle bin from all test particles that contribute at least 0.001 of the total flux in the bin, divided by the maximum contribution to the primary population peak bin for Orbit 64. The gray region is occupied by atoms making up the NIS He population. These atoms are practically absent on early orbits, like Orbit 57, and dominate on the orbits after the yearly peak of the signal, as Orbit 66. The right-hand column of panels presents geometric locations of the crossings of the orbits of atoms shown in the corresponding left-hand panels.  } \label{fig:rainbowPlot}
\end{figure}

The secondary population of interstellar gas has been directly sampled by IBEX-Lo in deuterium \citep{kubiak_etal:13a, rodriguez_etal:13a, rodriguez_etal:14a}, but the statistics is extremely poor and the interpretation is not certain. It was not detected in hydrogen \citep{schwadron_etal:13a}. If the Warm Breeze is indeed the heliospheric secondary helium population, its detection by IBEX-Lo implies that we have just opened a fascinating window to observe the physical state of plasma in the outer heliosheath almost directly, and with clear separation between the primary and secondary component. 

Our simulations show that it is possible to find direct links between the $6\degr$ spin angle intervals observed by IBEX-Lo and the geometric locations at which the observed atoms cross the 150~AU sphere around the Sun. Observing an individual $6\degr$ bin on a given orbit, we obtain information from a specific, well constrained region in the outer heliosheath. The atoms contributing to individual pixels cross the 150~AU sphere in well defined regions, which almost do not overlap. Images for three example orbits are shown in Fig.~\ref{fig:rainbowPlot}: one for the pure Breeze orbit 57, another for Orbit 60, on which the signal is composed of the Warm Breeze and NIS He contributions in comparable proportions, and the last one for Orbit 66, on which the Warm Breeze is barely visible at the fringes of the signal. This implies that IBEX-Lo is able to image $\sim 200\degr$ by $100\degr$  to $\sim 100\degr$ by $100\degr$ portions of the outer heliosheath, centered approximately on the upwind direction, shifting in space as the Earth moves around the Sun. The revisit time is one year, so, extending the IBEX mission to $\sim 10$ years would provide an opportunity to collect statistics sufficient to study the temporal behavior of the material in the outer heliosheath without the uncertainty of the region of the heliosphere from which the observed atoms originate. 

The grain of salt, however, is that presently we cannot state with certainty that the Warm Breeze is indeed the heliospheric secondary population. The observed signal from the Warm Breeze fits to a homogeneous Maxwellian population quite well. However, \citet{katushkina_izmodenov:10} demonstrated on the basis of  kinetic Monte Carlo modeling that even in the absence of interstellar magnetic field, the secondary component of interstellar hydrogen cannot be represented by a homogeneous Maxwellian population. We suspect that the same may be true for the secondary helium. Results of secondary helium simulation obtained in our paper show that both radial and angular gradients in the parameters of the secondary population are significant. They are not, however, included in our determination of the Warm Breeze parameters and we are not sure if this complex behavior of the secondary population gas in front of the heliosphere can be satisfactorily approximated by a homogeneous Maxwellian distribution function. The flow we actually observe seems consistent with the parent Maxwellian population with spatially homogeneous parameters. More tests are needed to see if a different, more complex distribution function better fits to the observations. Making such tests will be statistically warranted when the distribution of the IBEX-Lo background is better understood.

With the interpretation of the Warm Breeze as due to the secondary NIS He population not certain (even though the evidence in favor is strong), we have to think of alternatives. 

\subsubsection{Warm Breeze is due to elastic scattering of NIS He atoms on heliospheric ENAs in the outer heliosphere or beyond}

Another potential possibility for the source of the Warm Breeze would be elastic scattering of NIS He atoms on heliospheric particle populations. One of the reactions -- elastic scattering of NIS He on solar wind ions -- was addressed by \citet{gruntman:13a}, and shown to be insufficient in intensity to explain the observations. 

Potentially, however, a relatively small portion of NIS He atoms in front of the heliosphere might react by elastic scattering with Energetic Neutral Atoms (ENAs) created in the supersonic solar wind and in the inner heliosheath, running away from the heliosphere. Both H and He ENA are potentially important, H ENA being more abundant, but He ENA having much larger cross section for the reaction. The idea that heliospheric ENAs may heat the NIS gas in front of the heliosphere in the context of average plasma/gas parameters has been addressed in modern kinetic models of the heliosphere \citep[e.g.,][]{izmodenov_etal:09a}, but not in the context of creating a distinct population of NIS He atoms with thermal energies. 

Since the neutral gas in front of the heliosphere is collisionless, the scattering products should form a separate, distintct population, differing in general flow parameters both from the ambient interstellar gas and from the parent ENAs. %
Since the probability of the reaction would be relatively low, the process, to be efficient, would need to be one-step, i.e., a NIS He atom would have to lose approximately half of its momentum in the solar frame in a single collision. However, as shown in Appendix \ref{sec:appendixB}, 
the scattering is mostly forward, which is unfavorable for the creation of the Warm Breeze. 

\subsection{Warm Breeze is of extraheliospheric origin}

The idea that the heliosphere may be penetrated by a flow of neutral atoms from nearby astrophysical objects
was put forward by \citet{grzedzielski_etal:10b}, however in the context of IBEX ribbon, not neutral interstellar gas. \citet{frisch_etal:13a} suggested that 
the Sun may be traversing a region in the LIC 
which is turbulent and thus the interstellar matter surrounding the Sun may be inhomogeneous. Based on observations of interstellar absorption lines, \citet{frisch:94} and independently \citet{lallement_etal:95a}, \citet{wood_etal:00a}, \citet{redfield_linsky:00} hypothesized that the Sun may be very near the boundary of the LIC, perhaps within less than $\sim 10~000$~AU ($\sim 0.05$~pc). These considerations make us wonder whether the Warm Breeze might be a flow of interstellar matter not thermalized with the LIC material. 

If the Warm Breeze is a gust of neutral He relative to the LIC, its velocity relative to the LIC is equal to $\vec{v}_{WB-LIC} = \vec{v}_{WB} - \vec{v}_{LIC} = (v~=~13.6$~km/s, $\lambda = 93\degr$, $\phi = -4.5\degr$), where $\vec{v}_{LIC}$ is the LIC mean velocity vector that \citet{redfield_linsky:08a} obtained from the triangulation of Doppler shifts of interstellar matter absorption lines observed in the spectra of nearby stars \citep{crutcher:82a, bzowski:88}. This implies that the Warm Breeze material propagates through the LIC at $\sim 14$~km/s. 
As we show in Appendix \ref{sec:appendixB}, such flows can propagate without thermalization at distances of the order of several thousand AU, i.e., on spatial scales comparable to the distance between the Sun and the boundary of the LIC. Thus we consider a few possible scenarios that could generate such gusts in the immediate solar neighborhood. 

\subsubsection{Warm Breeze is due to LIC-G interaction, with the Sun close to the cloud-cloud boundary}

If the Sun is within $\sim$0.05~pc from the LIC boundary, one potentially could expect that a cloud-cloud or cloud-Local Bubble boundary layer may be created and the Sun embedded in it. Then, via a mechanism perhaps similar to the mechanism discussed by \citet{grzedzielski_etal:10b}, one could think of an inflow of non-thermalized neutral helium into the LIC. The velocity vector in the LIC reference frame would point from the nearby LIC boundary inward. The Colorado model of the LIC \citep{redfield_linsky:00} predicts that the LIC boundary is expected in the fourth quadrant of Galactic longitudes, so the direction of the Warm Breeze should be from this quadrant. But the velocity vector of the Warm Breeze in the LIC reference frame points towards $l \simeq 311\degr, b \simeq 32\degr$ in galactic coordinates, i.e., exactly in the opposite direction, which makes the hypothesis of the Warm Breeze coming from the LIC-G cloud interaction unlikely.

\subsubsection{Warm Breeze is a result of variations in the LIC plasma imparted to neutral helium}

In this hypothesis, the Warm Breeze is related to a possible non-uniformity of the LIC, i.e., temporal/spatial variations of plasma velocity/temperature/density. If fluctuations of  appropriate scale are present in the LIC ion density distribution, then He atoms created by (single) charge exchange neutralization of He$^+$ and/or (double) charge exchange neutralization of $\alpha$-particles in one place may be carried ballistically to another place and appear there as an additional population (gust), with kinematic characteristics different from the local neutral He population. Such a mechanism could be a viable explanation of the Warm Breeze if (1) fluctuations of suitable scale and magnitude do occur frequently enough in the LIC plasma, and (2) the effective ballistic range for the He atoms is sufficiently large.

Concerning (1), 
the existence of fluctuations can be deemed plausible based on what we know in general about (fluid type) waves in astrophysical plasmas. Suppose large amplitude, non-linear running waves, described by Riemann-type solutions for a gas with adiabatic (polytropic) relationships between pressure/temperature/density, propagate in the LIC. Then correlations between local fluid velocity/temperature/density will appear, with parcels of highest velocities corresponding to highest temperatures. Neutral He atoms transported ballistically from these parcels will be observed in neighboring areas as gusts of different velocities (in the LIC frame) and temperatures. 

A preliminary study of such correlations in the LIC was presented by Grzedzielski (2004) (also Grzedzielski et al., 2014, in preparation). As observational material, UV measurements of LIC Mg~II absorption lines in the spectra of 18 Hyades stars \citep{redfield_linsky:01a} were used. The lines-of-sight to these stars form a narrow beam in a direction almost opposite to the vector of Sun’s velocity in the LIC frame. The data show a correlation between the radial velocities (in the LIC frame) and spectral widths of absorption lines, with the correlation coefficient on the order of $>\sim$0.6. Such an effect could be explained if one assumes that a planar train of adiabatic, large amplitude Riemann waves propagates through the LIC. A numerical fit for the LIC with average $T = 7260$~K requires wave velocity amplitudes up to 1.5$\times$(average sound velocity) and a wave vector inclined at an angle $\zeta \sim 55^{\circ}$ to the Sun-apex line. 

To get a numerical example in the present context, we take the wave amplitude equal to average sound velocity in the LIC. Then the maximum temperature is $(1+ \frac{\gamma-1}{2})^2 \times$(average temperature), which leads to 12900~K for adiabatic exponent $\gamma$~=~5/3. With average sound speed $\simeq 10$~km/s, the parcel of maximum temperature will move in the solar frame towards the upwind hemisphere with a velocity of [(22~km/s~--~10~km/s~$\times \cos 55\degr)^2 + (10$~km/s~$\times \sin~55\degr^2]^{1\over2} = 18$~km/s at an angle $\arctan[(10$~km/s~$\times \sin 55\degr)/(22$~km/s$ -10$~km/s~$\times \cos 55\degr)]=27^{\circ}$ with respect to the Sun-apex line. To get a better agreement with the observed velocity of the Warm Breeze, $\zeta$ would have to be smaller, but the magnitude of velocity can be reconciled with the observed value and the Hyades direction was selected because it is the only we know of where observations towards many nearby stars are available, enabling an estimate as the one just presented. 

Concerning (2), analysis of elastic scattering and charge-exchange processes of He, provided in Appendix \ref{sec:appendixB}, indicates that the ballistic range of He atoms in the LIC can be on the order of $10^4$~AU. Therefore large amplitude waves in the LIC as tentatively suggested by analysis of the interstellar absorption lines in the direction of the Hyades could in principle produce gusts of atoms with kinematic characteristics different from the locally observed population, provided the wavelength is less than the mentioned range of $\sim$ 10$^4$~AU. On the other hand, very short waves,  besides being probably quickly damped, would smear out the effect. 

Thus the hypothesis that the Warm Breeze is evidence of waves of $\sim 10^4$~AU wavelength, propagating in the LIC plasma, may be plausible.

\section{Conclusions}

We analyzed observations of NIS He gas by IBEX-Lo from the 2010 observation seasons, adopting data points left out from previous analysis by \citet{bzowski_etal:12a}. We have discovered the Warm Breeze -- a new population of neutral helium in the heliosphere. The Warm Breeze is consistent with an inflow of a separate Maxwellian {population of neutral helium gas from $\sim 230\degr < 240\degr < \sim 250\degr$ in ecliptic longitude and $\sim 8\degr < 11\degr < \sim 18\degr$ in latitude, i.e., $\sim 19\degr$ to the side and $\sim 6\degr$ to the north of the inflow direction of neutral interstellar helium, at a speed of $\sim 7 <11 < 15$~km/s. It has a temperature of $\sim 7000 < \sim 15000 < \sim 21000$~K. Its density at the heliopause is $\sim 0.07 \pm 0.03$ of the density of the NIS He gas. The region of possible parameters in the 5D parameter space forms relatively narrow alleys, so other combination of parameters cannot be ruled out. Mild departures of the Warm Breeze distribution function at the heliopause from the perfect Maxwellian shape cannot be excluded, but remain one of the topics for future studies. Following the nautical terminology introduced to the heliospheric physics by \citet{mccomas_etal:13a}, we see a gust of the apparent wind, blowing from the starboard bow. 

The most likely explanation for the Warm Breeze may be the secondary population of NIS He atoms, created due to charge exchange between NIS He atoms and interstellar He$^+$ ions in the outer heliosheath, perhaps supplemented by elastic scattering of NIS He atoms on the heliosheath ion populations. The temperature, density, speed, and offset angle of the inflow direction qualitatively agree with model predictions. If this interpretation is true, observations of the IBEX-Lo pixels corresponding to the Warm Breeze signal, which we identify in this paper, offer an unprecedented opportunity of direct insight into the physical state of the outer heliosheath plasma. However, the macroscopic parameters of the secondary population, predicted by the heliospheric model we used, significantly vary in the region that ballistics dictates as the source for the Warm Breeze, which seems consistent with a homogeneous Maxwellian gas at 150~AU in front of the heliopause, so this interpretation cannot be regarded as certain. 

An alternative explanation for the Warm Breeze may be a gust of neutral He hurled through the LIC by a train of waves in the LIC plasma. Such gusts, as we show, may propagate for $\sim$10$^4$~AU before they dissipate and thermalize with the ambient gas, and the existence of the parent waves may be supported by observations of the correlation between the radial velocities and spectral widths of interstellar absorption lines towards the tightly grouped stars from the Hyades. If the latter hypothesis is true, the Warm Breeze and possible other interstellar gusts that may be present in IBEX data, but have not been sought for up to now, may offer an unexpected and exciting insight into microscale processes operating in the LIC.

\appendix
\section{Appendix: Simulation model for the secondary component of interstellar helium}
\label{sec:appendixA}
The global heliospheric model we used to model the secondary component of interstellar He is an extension of the Moscow Monte Carlo model, originally developed by \citet{baranov_malama:93} and expanded to include the interstellar magnetic field by \citet{izmodenov_etal:05a}. Effects of interstellar He$^+$ ions and solar wind alpha particles on the heliosphere were discussed by \citet{izmodenov_etal:03a}. 

Within the framework of the model used in the present paper, it is assumed that all charged components co-move and feature identical temperatures. To find the concentration of a separate charged component, He$^+$ in our case, one has to solve the continuity equation for the He$^+$ density $\rho_{He^+}$:
\begin{equation}
\frac{\partial \rho_{He^+}}{\partial t} + \nabla \cdot (\vec{v} \rho_{He^+})=q_{He^+},
\label{rownA1}
\end{equation}
assuming for He$^+$ a common local ambient plasma velocity $\vec{v}$. The right-hand side term $q_{He^+}$ is the source term, which can be calculated via solving a kinetic equation for the distribution function of the neutral He component $f_{He}(\vec{r},\vec{w}_{He},t)$ given by:
\begin{equation}
 \begin{aligned}
\frac{\partial f_{He}}{\partial t}+\vec{w}_{He}\cdot \frac{\partial f_{He}}{\partial \vec {r}} +\frac{\vec{F}}{m_{He}}\cdot \frac{\partial f_{He}}{\partial \vec{w}_{He}} = \\
= -f_{He}\int |\vec{w}_{He}-\vec{w}_{He^+}|\sigma^{He^+He}_{ex}f_{He^+}(\vec{r},\vec{w}_{He^+})d\vec{w}_{He^+} +\\
f_{He^+}(\vec{r},\vec{w}_{He})\int |\vec{w}^*_{He}-\vec{w}_{He}|   \sigma^{He^+He}_{ex}f_{He}(\vec{r},\vec{w}^*_{He})d\vec{w}^*_{He}-\beta f_{He}.
 \end{aligned}
\label{rownA2}
\end{equation}

In this equation, $\vec{w}_{He^+}$ and  $\vec{w}_{He}$ are individual velocity vectors for He$^+$ ions and He atoms, respectively; $\vec{F}$ is the solar gravity force acting on the He atoms; $\vec{r}$ is the radius vector of a particle; $\sigma^{He^+He}_{ex}$ is the relative velocity-dependent charge exchange cross section, adopted from \citet{phaneuf_etal:87}, following \citet{bzowski_etal:12a}; $\beta$ is the sum of the electron impact and photoionization rates; and $m_{He}$ the He atom mass. The source term $q_{He^+}$ in Eq.~\ref{rownA1} is given by the local ionization rate of neutral He via processes other than charge exchange: 
\begin{equation}
q_{He^+}=\beta m_{He} \int f_{He}(\vec{w}_{He})d\vec{w}_{He}=\beta m_{He} n_{He},
\label{rownA3}
\end{equation}
 
To close the system, we take the equation of state for plasma $p = (n_p + n_e + n_{He^+})\, k\, T$, where it is assumed that the temperature of  the He$^+$ component is equal to the ambient plasma temperature: $T=T_{He^+} $, and that the local He$^+$ plasma distribution function is given by
\begin{equation}
f_{He^+}(\vec{w}_{He^+})=n_{He^+}(\sqrt{\pi}u_{T,He^+})^{-3} \exp\left[-\frac{(\vec{w}_{He^+}-\vec{v})^2}{u^2_{T,He^+}}\right],
\label{rownA4}
\end{equation}
where $n_{He^+}$ is the He$^+$ ion concentration (number density), $u_{T,He^+}=(2kT/m_{He^+})^{1/2} $ is the thermal velocity of He$^+$ ions, $k$ -- the Boltzmann constant, and $m_{He^+}$ is the mass of He$^+$ ion.

In this calculation we neglect other potential charge exchange sources and sinks of neutral He because the combinations of relevant cross sections and abundances makes these reactions practically negligible \citep{bzowski_etal:12a, scherer_etal:14a}. Generally, one should add in the MHD equations the appropriate momentum and energy source terms related to the motion of He atoms. Since the modeling approach involves iterating MHD solutions for the plasma terms and Monte Carlo simulations for neutral atoms (see \citet{baranov_malama:93}), one would need to iterate the global solution both for H, which is the main driver, and for He. Since our modeling is just a qualitative reconnaissance of the potential secondary population of neutral He, we neglect the potential dynamical influence of neutral He on the global shape of the heliosphere and, consequently, we adopt the calculated sums of the charged components as fixed. With this, we can determine the distribution of helium atoms in ions in one integration process.

The boundary conditions adopted in the calculation are identical as adopted by \citet{izmodenov_etal:05a}. For the helium component in the LIC, we adopt $n_{He^+,LIC}=0.009$~cm$^{-3}$, $n_{He,LIC}=0.015$~cm$^{-3}$.
The distribution function of He atoms in the unperturbed LIC is similar to the distribution function of H and given by the Maxwell-Boltzmann formula defined in Eq.~\ref{rown5}.
The assumed concentration of He atoms in the unperturbed LIC is $n_{He,LIC}$, and the velocities and temperatures of the H and He components are adopted as: $\vec{v}_{He,LIC}=\vec{v}_{p,LIC}=\vec{v}_{LIC}$, $T_{He,LIC}=T_{p,LIC}=T_{LIC}$.

The model depends on the following nine parameters: (1) ionization degree in the LIC, (2) ratios of ram pressures, (3) ratios of temperatures, (4, 5) Mach numbers, (6, 7) ionization rates of H and He, and (8, 9) the cross the sections for charge exchange, respectively, for the LIC and solar wind. The solution of the problem was carried out by global iterations of the plasma and neutral kinetic models, similarly as proposed by \citet{baranov_etal:91}. 

\section{Appendix: Estimation of reasonable distance of non-thermalized neutral helium gusts in the LIC}
\label{sec:appendixB}
We estimate the spatial range of a neutral helium gust, propagating in the LIC, against elastic scattering. For the boundary conditions at the detector, we adopt the parameters of Warm Breeze as obtained in this paper, taken relative to the flow of the LIC material. We start the simulations assuming that the gust originates from a planar source and starts as a Maxwellian flow at some speed, density, and temperature different from the Warm Breeze parameters. This gust propagates through a homogenous LIC material, composed of protons, H and He atoms, and He$^+$ ions, and interacts with this material by elastic collisions. Collectively, this interaction results in continuous change of the gust speed, temperature, and density, which we track until they reach the Warm Breeze values. Arguably, it should also modify the local population of the LIC material, which, however, we leave out from the study as non-essential for its sole goal of assessing the range of the original population in the LIC.

The question of appropriate treatment of elastic scattering in the heliospheric environment was previously noticed by, e.g., \citet{izmodenov_etal:00a} and \citet{gruntman:13a}. The magnitude of the cross section for momentum transfer in elastic scattering of He atoms by ions and atoms in the LIC may suggest that the distance to the source of the Warm Breeze is approximately equal to the mean free path against collision, i.e., it should not exceed a few hundred AU. But the differential cross section in elastic scattering falls down very rapidly with the scattering angle, and in spite of the large mean scattering angle by protons, equal to $\sim$9$^{\circ}$ at the center-of-mass energy equal to 3~eV, more than half of scattering occur in an angle smaller than 1.5$^{\circ}$ (see Fig.~\ref{fig:integrated_xsev}). Thus a single scattering act does not eliminate the scattered He atom from the original population since the change of its momentum is not drastical. 

\setcounter{figure}{0}
\renewcommand{\thefigure}{B\arabic{figure}}

\begin{figure}
\epsscale{1.0}
\plotone{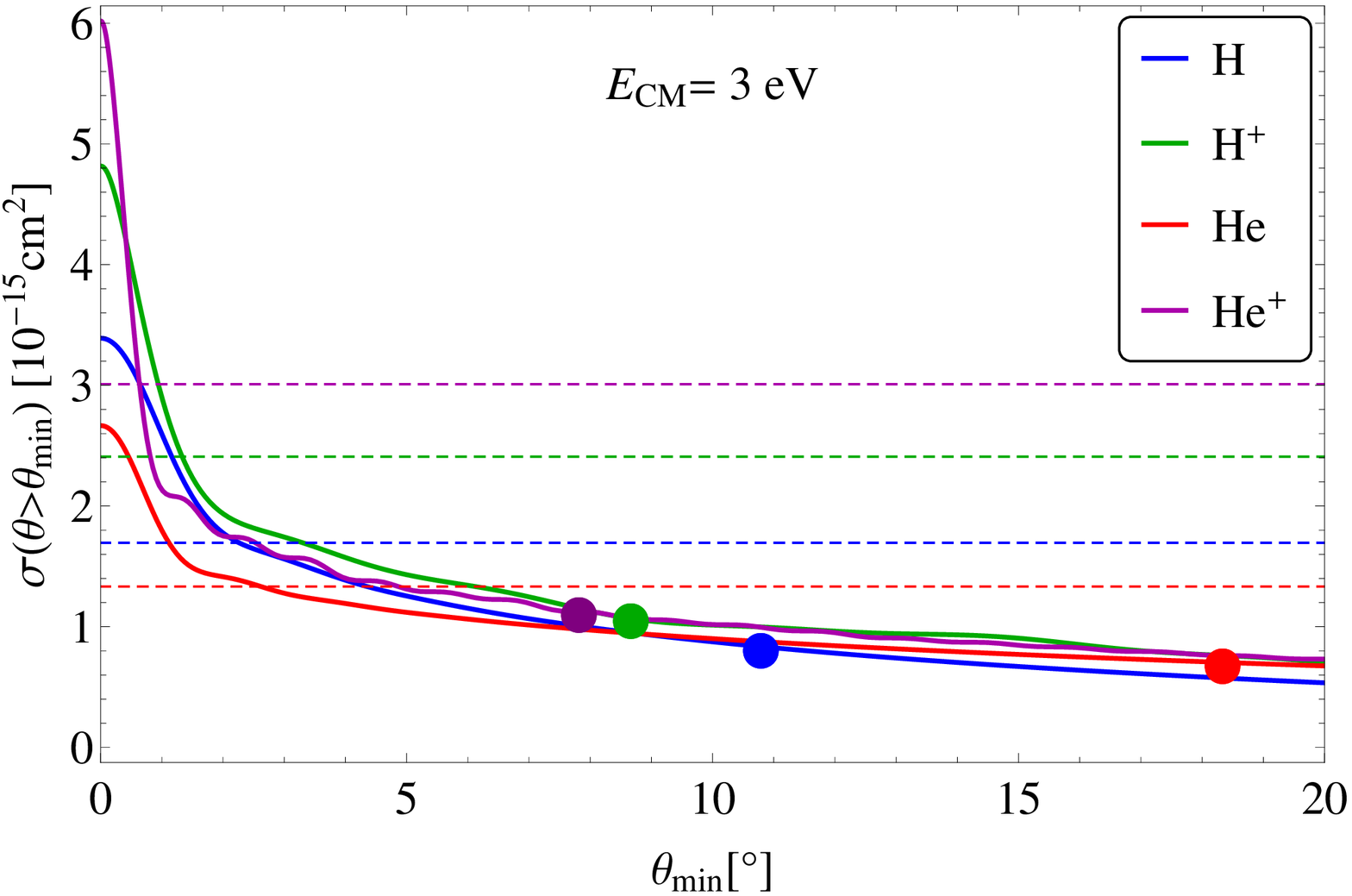}
\caption{Integrated elastic cross section of neutral He as a function of minimal scattering angle, taken as the lower limit of integration. The adopted collision energy in the center of mass reference system is 3~eV. Blue/green/red/purple lines denote the values for scattering of He on H / protons / He / He$^+$, respectively. Dashed horizontal lines present half of the integrated elastic cross sections in all scattering angles. Mean scattering angles are denoted by thick dots with colors corresponding to the scatterer. Note how strongly peaked is the cross section for the low (forward) scattering angles.}\label{fig:integrated_xsev}
\end{figure}

The evolution of Maxwellian distribution given in Eq.~(\ref{rown5}), where now $\vec{v}_B$ is relative velocity of NIS He and Warm Breeze in the LIC, was estimated using Monte Carlo simulation of elastic scattering. The simulation was performed for 10000 individual atoms, randomly chosen from the Maxwellian distribution. We assumed that the source of the Maxwellian distribution is a plane perpendicular to the velocity $\vec{v}_B$. The length of the simulation step was set at 10~AU. In each step, scattering occurs with a probability and angle given by the calculated elastic cross section on ions and atoms in the LIC. The simulation ends at a distance from the source 10 times longer than the mean free path for losses. The probabilities of survival were treated as weights in the calculations of population parameters at each distance from the source plane. Such treatment gives the same statistics at all distances from the source even if the survival probability for large distance is rather small.

\begin{figure}
\epsscale{1.0}
\plottwo{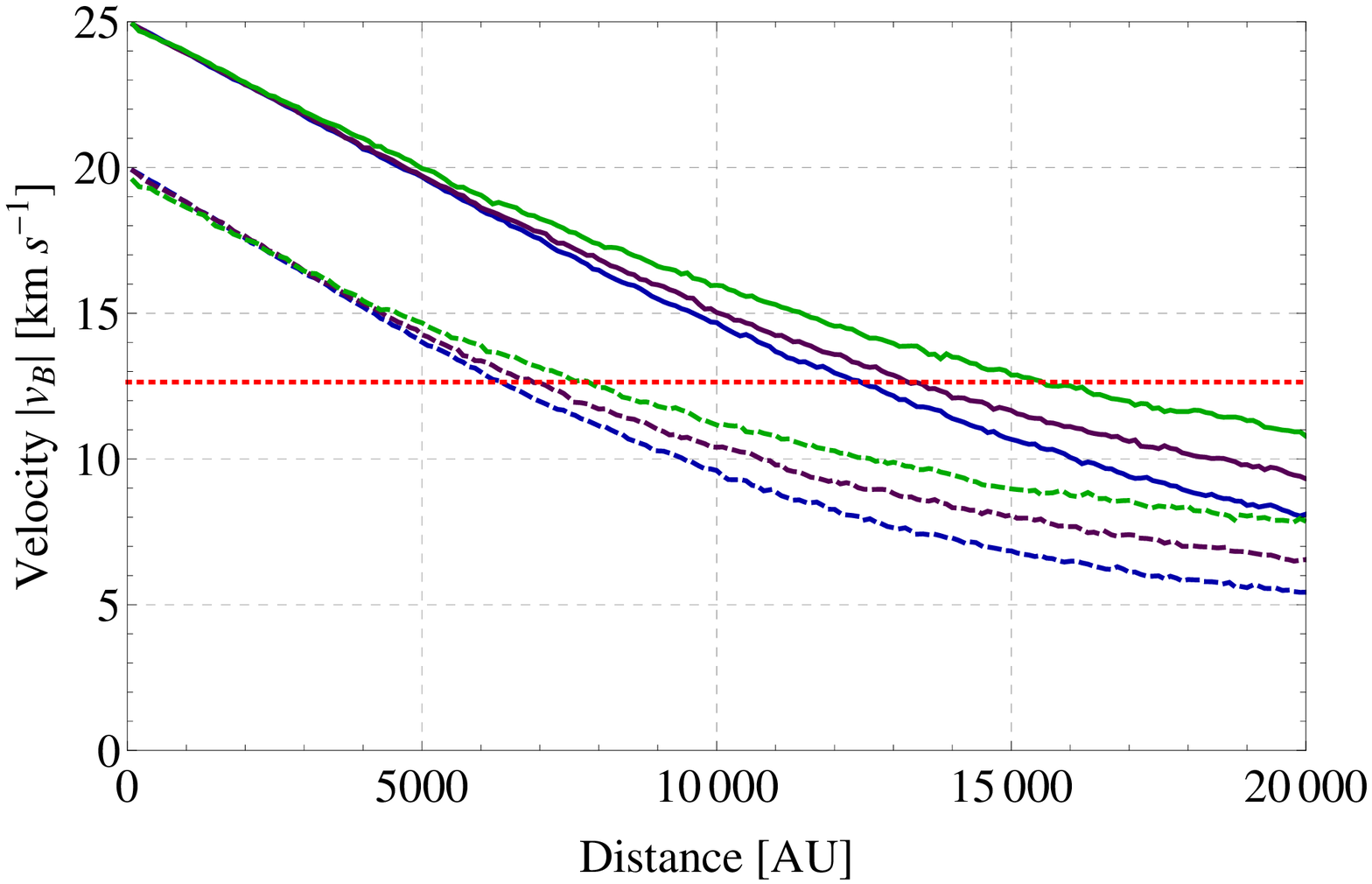}{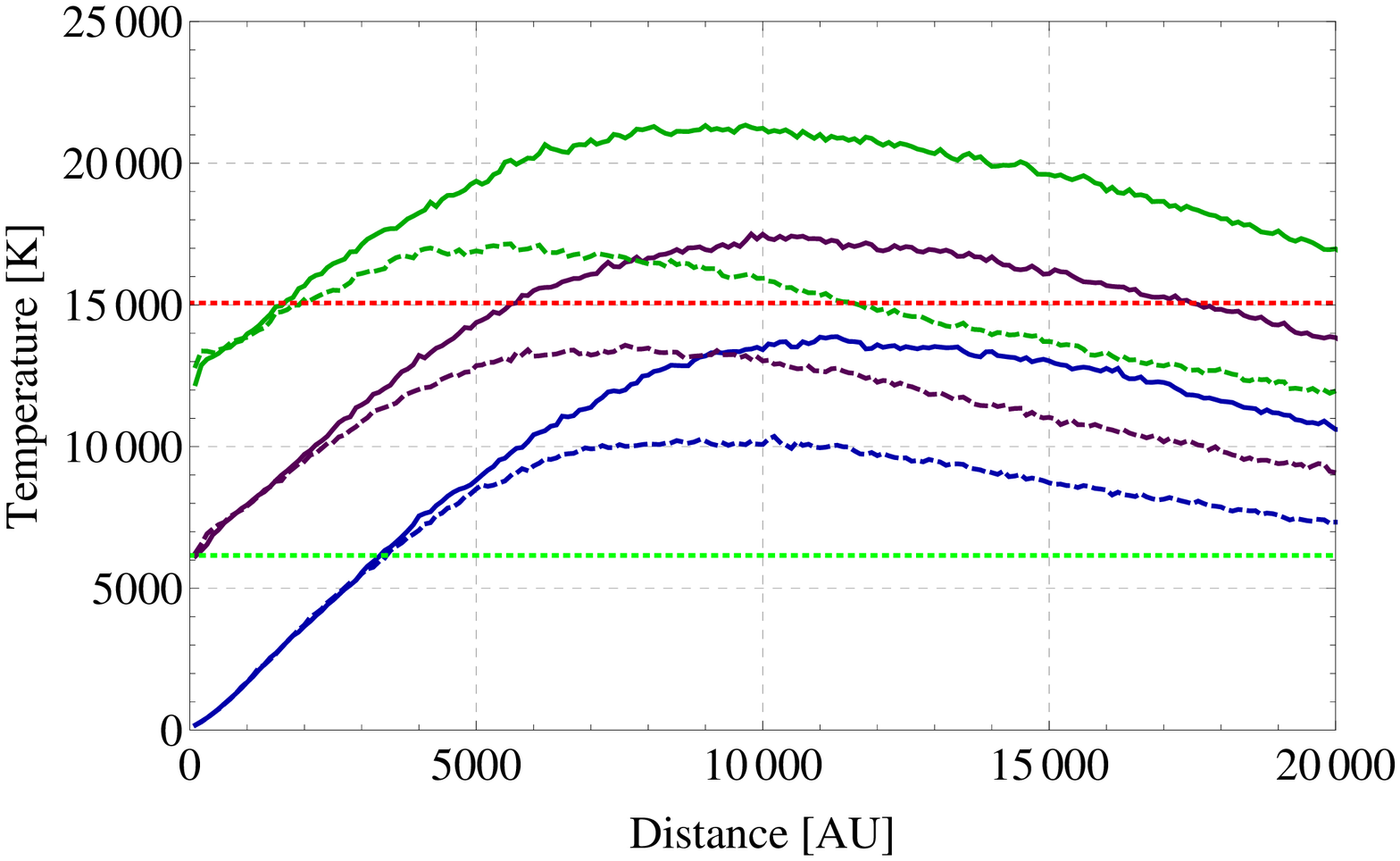}
\caption{Evolution of bulk velocity (left-hand panel) and temperature (right-hand panel) of hypothetic Maxwellian populations of neutral helium propagating through the LIC material and undergoing elastic scattering on the LIC He, He$^+$, H, H$^+$ ions and atoms, shown as a function of the distance from the source. The flow starts at six combinations of two initial bulk velocities (25 and 20~km/s, solid and broken lines, respectively) and three initial temperatures (100, 6000, and 12000~K, blue, purple, and green lines, respectively). The red line in the left-hand panel marks the flow velocity of the Warm Breeze relative to LIC. The red line in the right-hand panel marks the temperature of the Warm Breeze and the green line is for the temperature of the LIC. Full dissipation and thermalization of these hypothetic populations occurs when the bulk velocity relative to the ambient material disappears and the temperature is reduced to the temperature of the ambient gas.}.\label{fig:vb}
\end{figure}

The elastic scattering cross sections of neutral He atoms with various atoms and ions were calculated using the semi-classical JWKB approximation, as presented with details, e.g., by \citet{nitz_etal:87}. Potentials for the interactions were taken from analytic formulae for H-He \citep{gao_etal:89}, H$^+$-He \citep{wolniewicz:65, helbig_etal:70}, He-He \citep{ceperley_partridge:86} and He-He$^+$ \citep{barata_conde:10}. As a check for the procedure, we compared our results of the H$^+$-He differential cross section with those calculated by \citet{krstic_schultz:99}. For an energy in the center-of-mass system larger than 0.5 eV, the obtained integrated cross section and momentum transfer cross sections differ less than 10\%. Small differences in the positions of interference maximum that we have noticed are not important for our analysis. 

\begin{figure}
\epsscale{1.0}
\plotone{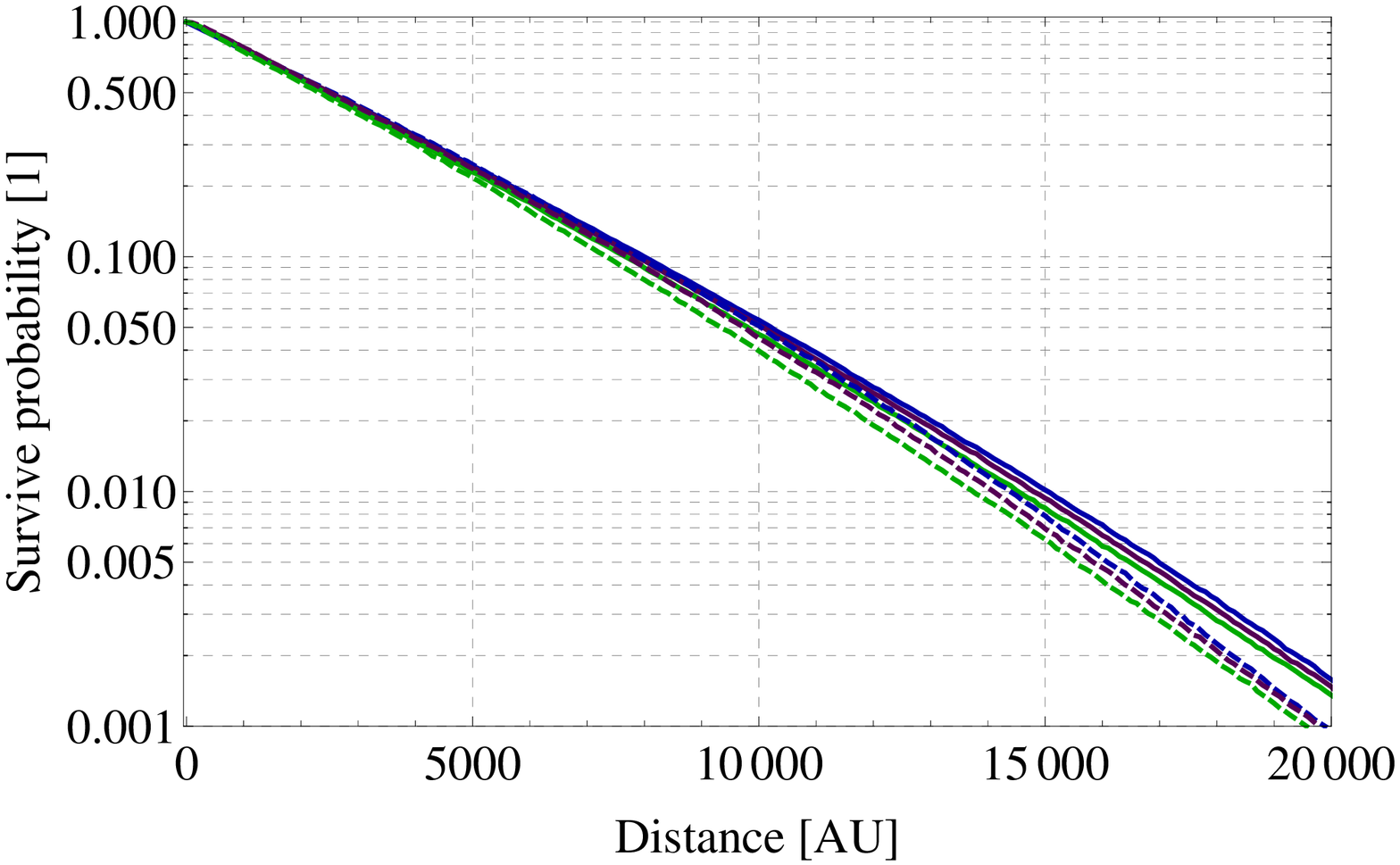}
\caption{Survival probabilities of He atoms against charge exchange with protons in the LIC for the six combinations of speeds and temperatures, as presented in Fig.~\ref{fig:vb}. The survival probabilities (corresponding to the attenuation of the beam as it propagates through the LIC) differ relatively little as a function of initial parameters and suggest that the original beam will reduce in its density to $\sim$7\% of the original density at $\sim10^4$~AU from the source.}\label{fig:probsurv}
\end{figure}

In addition to scattering, the He atoms from the gust are being lost because of ionization, predominantly by charge exchange with the ambient plasma ions. The mean free path for losses was determined based on the cross section for charge-exchange with He$^+$, which is orders of magnitude larger than the cross section for charge exchange with protons. The cross section in the energy range 1 -- 10~eV varies from 2.3~$\times$~10$^{-15}$~cm$^2$ to 1.7~$\times$~10$^{-15}$~cm$^2$ \citep{barnett_etal:90}. The ionic state of interstellar He in the LIC is thought to be 0.611~He, 0.385~He$^+$ and 0.00436~He$^{++}$ \citep{slavin_frisch:08a}. For NIS He density equal to 0.015~cm$^{-3}$ \citep{witte:04, gloeckler_etal:04a, mobius_etal:04a}, the mean free path for charge-exchange varies in the range 3000~AU -- 4000~AU. For simplicity, in our analysis we adopted it as 3500~AU. 

The simulations start at two different values of velocity $v_B$: 25~km/s and 20~km/s, and three temperatures 100~K, 6000~K, and 12~000~K. The evolution of $v_B$ and temperature as a function of distance from the source for the six combinations of starting parameters is presented in Fig.~\ref{fig:vb}. Starting with velocity 20 (25)~km/s, the gust is monotonically slowed down to 13~km/s at a distance $\sim$7000 (14000)~AU. The evolution of temperature is more complex due to randomization of velocity $v_B$ and thermalization. The first of them should be so effective that it should give increase of temperature of several thousand Kelvin. The temperature initially raises, reaches the broad peak value at $\sim$10000~AU from the source regardless of initial condition, and gradually falls down to the ambient LIC value, being thermalized at a distance of several dozen thousand AU (not shown in the figure).

Taking into account charge-exchange with He$^+$, the probability of reaching a distance of 10000~AU for a Warm Breeze atom is equal to $\sim$5\% (Fig.~\ref{fig:probsurv}). It suggests that for the Warm Breeze being observable as a few percent admixture to the LIC matter, the densities of NIS He and the Warm Breeze at the source should be comparable. 

Given these results, it is feasible to consider sources of non-thermalized flows in the LIC at distances of a few thousand AU from the Sun, which could be seen as populations of neutral He gas separate from the predominant NIS He population. These distances are considerably larger than the naively assumed limit of the mean free path against scattering or charge exchange.

\section*{Acknowledgment}
The research portion carried out at SRC PAS was supported by the Polish National Science Center grant 2012-06-M-ST9-00455. The work in the US was carried out under the IBEX mission, which is part of NASA’s Explorer program, under contract NNG05EC85 and grant NNX10AC44G. The work of D.A. and V.I. was supported by Russian Foundation of Basic Reserarch (grant 14-02-00746). Part of this research was carried out in the framework of International Space Science Institute Working Team 223.

\bibliographystyle{plainnat}
\bibliography{iplbib}{}

\end{document}